\documentclass[twocolumn,prb,amsmath,showpacs]{revtex4}

\usepackage{multirow}
\usepackage{graphicx}
\usepackage{subfigure}

\newcommand{\re}{\mathrm{Re}}

\newcommand{\w}{\omega}
\newcommand{\bw}{\bar{\omega}}
\newcommand{\bq}{\bar{q}}

\newcommand{\vk}{\mathbf{k}}
\newcommand{\vv}{\mathbf{v}}
\newcommand{\vq}{\mathbf{q}}
\newcommand{\vQ}{\mathbf{Q}}
\renewcommand{\vr}{\mathbf{r}}
\newcommand{\vM}{\mathbf{M}}
\newcommand{\vH}{\mathbf{H}}
\newcommand{\vB}{\mathbf{B}}

\newcommand{\cD}{\mathcal{D}}
\newcommand{\cH}{\mathcal{H}}
\newcommand{\cF}{\mathcal{F}}
\newcommand{\cO}{\mathcal{O}}
\newcommand{\cY}{\mathcal{Y}}
\newcommand{\cZ}{\mathcal{Z}}

\newcommand{\tEta}{\tilde{\eta}}
\newcommand{\tChi}{\tilde{\chi}}

\newcommand{\cecoin}{\mathrm{CeCoIn}_5}

\begin{document}

\title{Pauli-Limited Superconductivity with Classical Magnetic Fluctuations}


\author{Robert Beaird}
\author{Anton B. Vorontsov}
\altaffiliation{Present address: Department of Physics, Montana State University, Bozeman, MT 59717-3840, USA}
\author{Ilya Vekhter}
\affiliation{Department of Physics and Astronomy, Louisiana State University, Baton Rouge, LA, 70803-4001, USA}

\date{\today}

\begin{abstract}
We examine the effect of classical magnetic fluctuations on the phase diagram of paramagnetically-limited two-dimensional superconductors under a Zeeman magnetic field. We derive the free energy expansion in powers of the superconducting order parameter and analyze the character of the normal-superconducting transition. While the transition is of the second order for all temperatures in the absence of magnetic fluctuations, we find that proximity to magnetism drives both the transition into the uniform state and that into the modulated (Fulde-Ferrell-Larkin-Ovchinnikov, FFLO) state to first order at intermediate temperatures. We compute the thermodynamic signatures of the normal-superconducting transition along the upper critical field.
\end{abstract}

%\keywords{}
\maketitle

%%%%%%%%%%%%%%%%%%%%%%%   Introduction   %%%%%%%%%%%%%%%%%%%%%
\section{Introduction}

Since the discovery of the Meissner effect, there has been ongoing study into the interplay of superconductivity with magnetic fields. In type II superconductors, supercurrents, which arise from the coupling of the Cooper pair momenta to the vector potential of the magnetic field, cause the system to return to the normal state at the orbital-limited upper critical field $H_{c2,orb}$. If, however, the dominant coupling of spin-singlet superconducting electrons to the field is via the Zeeman effect, the transition occurs at the Pauli-limited field $H_P$ where the condensation energy equals the gain in energy due to spin polarization of the two electrons in a Cooper pair. In addition, in pure systems in the vicinity of this field, the superconducting state can exhibit spatial-modulation of the type predicted by Fulde and Ferrell, and Larkin and Ovchinnikov (FFLO) at low temperature and high magnetic fields\cite{FF,LO}.

No clear examples of FFLO superconductivity have been found; however, early experiments \cite{BianchiFFLO:2003, CapanThermalCond:2004, MicleaPressure:2006, MitrovicSpinSuscCeCoIn5:2006}  on the layered\cite{Petrovic:2001}, heavy-fermion superconductor CeCoIn$_5$, tentatively identified the low-temperature, high-field (LTHF) superconducting phase as a possible realization of the FFLO state. The heavy mass and large value of the ratio between the estimated orbital critical field and the Pauli-limiting field $\sqrt{2}H_{c2,orb}/H_P\simeq 3.5$, \cite{BianchiFOTCeCoIn:2002} suggest strong paramagnetic limiting. It was also proposed that the unusual field dependence of the vortex lattice form factor of CeCoIn$_5$ is due to Pauli-limiting effects\cite{IchiokaVortexPauli:2007,BianchiKenzelmannSCvortices:2008}.

Some experimental features of the transition into the LTHF phase are not fit by the established theories of Pauli-limiting. For example, it has been established experimentally\cite{BianchiFFLO:2003} that the transition from the normal (N) to superconducting (SC) state in CeCoIn$_5$ is first order into low-field state ($T<T_{FFLO}$), and remains first order at low temperatures up to $T_0>T_{FFLO}$. This is in sharp contrast to the conventional theory in two dimensions (2D) that finds a second order N-SC transition along the entire critical field line $B_{c}(T)$ \cite{BurkhardtFermiLiquidFFLO:1994,AntonFFLO2D:2005}, and motivates our current study.

Under purely Zeeman field, $B$, when the electron spins couple to the field but the orbital coupling to the vector potential is irrelevant, at low temperatures the N-SC transition is into an inhomogeneous state \cite{FF,LO}.  In most cases the amplitude modulated state, $\Delta(\vr)~=~\Delta_0~\sin(Qx)$ (LO state), is favored compared to the purely phase-modulated FF state, $\Delta(\vr)~=~\Delta_0~\mathrm{e}^{iQx}$. In the conventional analysis for isotropic, $s$-wave, superconductors the FFLO transition is 2nd order in 2D and 1st order in 3D \cite{BurkhardtFermiLiquidFFLO:1994,BuzdinFFLO:1997,CombescotMoraFFLO3D:2005,MatsuoFFLO3D:1998}.
In superconductors with nodes, such as $d$-wave CeCoIn$_5$\cite{Petrovic:2001, AokiDxyGapCeCoIn5:2005,
IzawaGapNodes:2001, MovshovichUncSC:2001}, the FFLO transition is 2nd order in both 2D\cite{AntonFFLO2D:2005} and 3D\cite{SamokhinFFLO:1997}. The transition to a combined vortex and LO state is also expected to be second order
\cite{GruenbergType2FFLO:1966}.

Theoretically, under several conditions the transition from the normal to the SC state may become first order. For purely Zeeman coupling this can happen a) due to strong Fermi-liquid enhancement of the magnetic susceptibility~\cite{BurkhardtFermiLiquidFFLO:1994,
AntonFFLO2D:2005}, or b) due to impurity scattering in the
resonant limit \cite{AntonDirtyDwave:2008}. In the presence of
both paramagnetic and orbital effects it was argued in
Refs.~[\onlinecite{AdachiIkeda:2003,HouzetMineev:2006}] that the transition may also become first order in an intermediate
temperature range.

In this paper we show that in systems with enhanced magnetic susceptibility (as compared to the dimensionless Pauli susceptibility for typical metals $\chi_{P}\approx10^{-6}$), such as some heavy fermion materials, first order N-SC transitions under purely Zeeman magnetic field may naturally emerge over a part of the phase diagram. Importantly, the transition is first order both for the uniform and for the FFLO state over part of the temperature range. We consider the critical field $B_{c}(T)$ and discuss how the magnetic fluctuations affect the order and thermodynamics of the transition.

The rationale for inclusion of such fluctuations is as follows. Experiments convincingly show that CeCoIn$_5$ is in proximity to a magnetically ordered state\cite{Sidorov:2002, Tayama:2002, BianchiAFMQCP:2003, PaglioneFieldIndQcpCeCoIn5:2003, KnebelPressureRhCoCompare:2004, PhamCdDoping:2006, ParkHiddenMagOrder:2006, MicleaPressure:2006, NicklasAFMflucts:2007, NicklasCdDopedCeCoIn5:2007, OhiraKawamuraAfmCeRhCoIn5:2007, YoungFieldInducedMag:2007, KenzelmannSCandMagOrder:2008}. In CeCoIn$_5$ the $f$-electron spins are not fully Kondo screened by the onset of the superconducting order~\cite{Petrovic:2001, Tayama:2002}, and the entropy of the remaining spin fluctuations is released at the superconducting transition\cite{Petrovic:2001, BianchiFFLO:2003}. The specific heat jump upon entering the superconducting state at $T_\mathrm{c}$ in zero field is $\Delta C/\gamma T_\mathrm{c} \sim 4.5$ (where $\gamma$ is the Sommerfeld coeffient), more than three times the $s$-wave BCS value of $1.43$ \cite{Petrovic:2001}. Guided by this observation, Kos, Martin, and Varma \cite{KMV:2003} considered a Ginzburg-Landau model of competition between thermal (Gaussian) magnetic fluctuations and superconductivity, and were able to explain the large specific heat jump at $T_c(B=0)$.

We employ similar methods, with more microscopic considerations, to include an applied magnetic field. To simplify our analysis, we consider only the Zeeman coupling since it is largely responsible for the salient features in the phase diagram (\emph{e.g.}, the first order transition). We explain our results in context of experiment for CeCoIn$_5$ and emphasize that our approach is generally applicable to Pauli-limited systems with thermal magnetic fluctuations.

The rest of the paper is organized as follows. In Sec.~II we review the physics of superconductivity in the paramagnetic limit and our methodology of determining the transition line. In Sec.~III we extend the formalism to account for the magnetic fluctuations, and use this method to obtain the results presented and discussed in Sec.~IV. We conclude by placing our results in the context of experiment and theory on Pauli-limited superconducting systems.

%%%%%%%%%%%%%%%%%%%   Theoretical Methods   %%%%%%%%%%%%%%%%%

\section{Superconductivity in the Paramagnetic Limit}
\subsection{Model and approach}

In the paramagnetic limit the second order transition into the
uniform superconducting state, $\Delta(\vr)=\Delta_0$, becomes first order below a characteristic temperature $T_P \simeq 0.56 T_{c0}$ where $T_{c0}$ is the transition temperature in zero field \cite{Maki:1964}. At $T=0$ superconductivity is destroyed when the energy of the polarized normal state equals the superconducting condensation energy (the Clogston\cite{Clogston:1962}-Chandrasekhar\cite{Chandrasekhar:2004} limit). This occurs at the Pauli field $H_P=\Delta/(\sqrt{2}\mu)$, where $\mu=g\mu_B/2$ is the electron magnetic moment, $\mu_B$ is the Bohr magneton, and $g$ is the conduction electron $g$-factor. The
microscopic pairbreaking occurs as the Zeeman field increases the energy of the spin-singlet with respect to spin-polarized $s=1$. An alternative to this uniform superconductivity is the pairing of the electrons with opposite spins and the same energies, which now have momenta differing by $Q\sim \mu B/(\hbar v_F)$, where $v_F$ is the Fermi velocity. The finite center of mass momentum of the Cooper pairs leads to a spatial modulation of the order parameter~\cite{FF,LO} and allows superconductivity to survive at fields above the Clogston-Chandrasekhar limit.

The exact structure of the modulated state in $s$-wave systems is still not well established. Generally the amplitude-modulated LO state is lower in energy than the current-carrying FF state~\cite{BuzdinFFLO:1997}. In the absence of spin-orbit coupling, the direction of $\mathbf{Q}$ in real space can be chosen arbitrarily, and a superposition of plane wave modulations along different directions may yield yet lower energy~\cite{ShimaharaSuperpositionFFLO:1998, CombescotMoraLowTemp2DCascade:2005}. In systems with unconventional, such as $d$-wave~\cite{SamokhinFFLO:1997, MakiSineLikeDwave:1996, AntonFFLO2D:2005,
MatsudaShimaharaReview:2007}, gap symmetry the modulation is
preferentially along either the nodal or anti-nodal orientation, depending on both the temperature and the purity of the sample\cite{AntonFFLO2D:2005,AntonDirtyDwave:2008}, with the LO state always more advantageous.

Analysis of the FFLO states is often carried out within the
Ginzburg-Landau theory, expanding the free energy of the
superconducting state in both the amplitude and the gradient of the order parameter~\cite{BuzdinFFLO:1997}. Such an expansion is justified in the immediate vicinity of $T_P$, but its region of validity is very narrow. The modulation wave vector increases rapidly along $T_c(B)<T_P$ and becomes comparable
to the inverse of the SC coherence length, $\xi_0^{-1}=\left[\hbar v_F/2\pi T_{c0}\right]^{-1}$, rendering the gradient expansion invalid. Previously we reported the results of a brief analysis of the N-SC transition for a superconductor with magnetic fluctuations under Zeeman field using such a gradient expansion~\cite{BeairdFluctsGrad:2008}. The results were suggestive of the onset of the first order transition near $T_P$. The limitations of the gradient expansion prevented us from reaching detailed conclusions, and motivated our present work. Below we expand the free energy only in powers of the order parameter and retain the full wave-vector dependence of the expansion coefficients, thus removing the deficiencies of the gradient expansion and presenting a theory valid down to low temperatures. This allows us to analyze the details of the phase diagram not accessible with the gradient expansion.

We begin with the mean-field Hamiltonian
\begin{equation}
\begin{split}
\cH =& \sum_{\vk,\sigma}\epsilon_{\vk,\sigma}
c_{\vk,\sigma}^\dagger
c_{\vk,\sigma}+\frac{1}{|\lambda|}\sum_{\vq}|\Delta_\vq|^2\\
 &-\sum_{\vq,\vk}\cY(\hat{\vk})\left(\Delta_\vq c_{\vk+\vq,+}^\dagger
c_{-\vk,-}^\dagger + h.c.\right)\,,
\end{split}
\label{Hmf}
\end{equation}
where $\sigma=\pm$ denotes the orientation of the electron with  the spin along/opposite to the field direction, $\epsilon_{\vk}$ is the band energy measured with respect to the chemical potential, and $\epsilon_{\vk,\sigma} = \epsilon_\vk +\sigma \mu B$. In Eq.~(\ref{Hmf}), $|\lambda|$ is the strength of the pairing interaction, $\cY(\hat{\vk})$ is a normalized basis function that transforms according to an irreducible representation of the crystal point group and describes the gap symmetry, and $\hat{\vk}$ denotes position on the Fermi surface (FS).

We assume, for simplicity, a
separable pairing interaction, so that the spin-singlet order
parameter is $\psi(\vk,\vq)=\cY(\hat{\vk})\Delta_\vq$, with the amplitude $\Delta_\vq$ self-consistently determined from
\begin{equation}
\label{deltaq}
\Delta_\vq=
-|\lambda|\sum_\vk\cY(\hat{\vk})
\left<c_{\vk+\vq,+}c_{-\vk,-}\right>\,
\end{equation}
where $\langle\cdots\rangle$ indicates thermal average. Uniform superconducting states have the single non-vanishing Fourier component with $\vq=0$, while modulated states correspond to one or more components with $\vq\neq 0$. Since Eq.(\ref{deltaq}) has to minimize the free energy, it determines, at the mean field level, the Landau expansion of the free energy density $F_{L}$ in powers of $\Delta_{\vq}$,
\begin{equation}\label{gen glf}
\begin{split}
    F_{L} &= \sum_{ \{\vq_i\}}\widetilde{\alpha}_{\vq_i} |\Delta_{\vq_i}|^2 \\
    & +\sum_{ \{\vq_i\}}\widetilde{\gamma}_{\vq_1,\ldots,\vq_4}
    \Delta_{\vq_1}\Delta^*_{\vq_2}\Delta_{\vq_3}\Delta^*_{\vq_4}\delta_{\vq_1+\vq_3,\vq_2+\vq_4}\\
    & +\sum_{ \{\vq_i\}}\widetilde{\nu}_{\vq_1,\ldots,\vq_6}
    \Delta_{\vq_1}\Delta^*_{\vq_2}\Delta_{\vq_3}\Delta^*_{\vq_4}\Delta_{\vq_5}\Delta^*_{\vq_6}\\
    & \hspace{2cm}\times
    \delta_{\vq_1+\vq_3+\vq_5,\vq_2+\vq_4+\vq_6}\,.
\end{split}
\end{equation}
The coefficients of this expansion are combinations of the normal state Green's functions as described in Appendix~\ref{GLF derivation}. The summation over $\{\vq_i\}$ in Eq.~\eqref{gen glf} includes all possible combinations of the allowed Fourier components of $\Delta (\vr)$: $\vq_i=0$ for a uniform gap amplitude, single mode $\vq_i=\vQ$ for the FF modulation, and $\vq_i\in\{\vQ,-\vQ\}$ for the LO phase. In the following we restrict ourselves to the comparison of the free energies of these three phases, finding the one most energetically favorable and the corresponding wave vector $\vQ$.

Based on the observation of the quasi two-dimensional Fermi surface in the 115 family~\cite{HallFSCeCoIn5:2001, SettaiFSCeCoIn5:2001, HagaFermiSurfaceCeIrIn5:2001, ShishidoFSdHvAPressure:2003}, we use a model of a 2D circular Fermi surface. We use the azimuthal angle, $\theta$, to parameterize the position on the Fermi surface, and we choose $\cY(\hat{\vk})\equiv\cY(\theta)=1$ and  $\cY(\theta)=\sqrt 2 \cos 2\theta$ for $s$- and $d$-wave gaps, respectively.

We determine the phase transition line $B_c(T)$ by finding, at a given temperature, $T$, the highest $B_c$ of the three phases we compare. In each phase we find $B_c(T)=\max(B_c(T,\mathbf{q}))$ by unrestricted maximization with respect to the modulation wave vector. We introduce the dimensionless energy density, $f=F_{L}/N_F T_{c0}^2$, where $N_F$ is the 2D normal state density of states at the Fermi level.  We also introduce the dimensionless amplitude $\delta_0=\Delta_0/T_{c0}$ where $\Delta_0$ is the SC gap amplitude and $T_{c0}$ is the mean-field transition temperature at $B=0$ in the absence of magnetic fluctuations. We set $k_B=\hbar=1$ throughout the paper.  The reduced temperature and magnetic field are given by $t=T/T_{c0}$ and $b=\mu B/(2\pi T_{c0})$ respectively.

%%%%%%%%%%%%%%%%%%%%%%%%%%%%%%%%%%%%%%%%%%%%%%%%%%%%%%%%%%%%%%%%%%%%%%%%%%%%%%%%%%%%%%%%%
\subsection{Uniform superconducting state}%: $\Delta(x)=\Delta_0$}

For the uniform state, $\Delta_\vq = \Delta_0\delta_{\vq,0}$, we find
\begin{equation}
f_u(T,B) = \alpha_u|\delta_0|^2+\gamma_u|\delta_0|^4
+\nu_u|\delta_0|^6\,,
\end{equation}
with the coefficients determined from Eqs.~\eqref{alpha
general}-\eqref{nu general},
\begin{subequations}\label{eq:GLu bare}
\begin{eqnarray}
\label{alpha u bare}
\alpha_u&=&\ln\left(t\right)+\re\left[\Psi\left(\frac{1}{2}+i\frac{b}{t}\right)\right]-\Psi\left(\frac{1}{2}\right),
\\
\label{gamma u bare}
\gamma_u&=&-\frac{1}{8}\frac{\left<|\cY(\theta)|^4\right>_{FS}}{(2\pi
t)^2}\re\left[\Psi^{(2)}\left(\frac{1}{2}+i\frac{b}{t}\right)\right],
\\
\label{nu u bare}
\nu_u&=&\frac{1}{192}\frac{\left<|\cY(\theta)|^6\right>_{FS}}{(2\pi
t)^4}\re\left[\Psi^{(4)}\left(\frac{1}{2}+i\frac{b}{t}\right)\right]\,.
\end{eqnarray}
\end{subequations}
Here $\Psi$ ($\Psi^{(n)}$) is the digamma ($n$th order polygamma) function, and $\langle\cdots\rangle_{FS}=\int d\theta/(2\pi)$. For the $s$- and $d$-wave symmetries of the gap, Eq.~\eqref{basis fxns}, our coefficients agree with those in
Refs.~[\onlinecite{BuzdinFFLO:1997,SamokhinFFLO:1997}].

%%%%%%%%%%%%%%%%%%%%%%%%%%%%%%%%%%%%%%%%%%%%%%%%%%%%%
\subsection{FF state}

For the spatially-inhomogeneous superconducting state, the
coefficients in Eq.~\eqref{gen glf}, depend on the direction of modulation. Since the modulation wave vector $Q\sim \xi_0^{-1}\ll k_F$, for two particles at locations $\theta$ and $\pi+\theta$ on the Fermi surface, there is an energy mismatch  $\vv_F\cdot \vQ = v_FQ\cos(\theta-\theta_Q)$, where $\theta_Q$ is the modulation direction with respect to the crystalline $a$ axis. This energy mismatch enters in Eq.~\eqref{alpha general} with $\vq_i=\vQ$.

Recall that the polygamma functions in Eq.~\eqref{eq:GLu bare} originate from the summation over Matsubara frequencies, and that their argument is determined by the energy mismatch of the particles in the Green's functions in Eqs.~\eqref{alpha general}-\eqref{nu general}. Consequently, the coefficients of the free energy expansion in the FF state are given by the same polygamma functions as for the uniform case, Eq.~\eqref{eq:GLu bare}, but with the arguments reflecting the energy difference $\mu B+\vv_F\cdot \vQ$. Hence in the expansion $f_{FF}(T,B) = \alpha_{FF}|\delta_0|^2+\gamma_{FF}|\delta_0|^4
+\nu_{FF}|\delta_0|^6$, we find
\begin{equation}\label{alpha FF bare}
\begin{split}
\alpha_{FF}&=\ln\left(t\right)-\Psi\left(\frac{1}{2}\right)+\\
&\re\hspace{-0.5mm}\left<|\cY(\theta)|^2\Psi\left(\frac{1}{2}+i\frac{b
+ \bar{q}}{t}\right)\right>_{FS}\,,
\end{split}
\end{equation}
where $\bar{q}=q \cos(\theta-\theta_q)$ and $q =\xi_0 Q/2$.
Similarly, $\gamma_{FF}$  and $\nu_{FF}$ are given by expressions identical to Eqs.~\eqref{gamma u bare} and \eqref{nu u bare} under the replacement
$b\rightarrow b+\bar{q}$ and averaging both the digamma functions and the basis functions $\cY(\theta)$ together over the Fermi surface.

It follows that for any anisotropic superconductor the direction of the modulation and the shape of the gap cannot be separated. For a two-dimensional $d$-wave superconductor that we consider, the modulation along the nodal/antinodal direction is preferred in a pure material above/below
$T\simeq0.06T_{c0}$\cite{AntonFFLO2D:2005,YangSondhi:1998,ShimaharaSuperpositionFFLO:1998},
although as the impurity scattering is increased modulation along a node becomes favorable even for $T<0.06T_{c0}$\cite{AntonDirtyDwave:2008}. Therefore, below we focus on the modulation along the gap nodes.

%%%%%%%%%%%%%%%%%%%%%%%%%%%%%%%%%%%%%%%%%%%%%%%%%%%%%%%%%%%%%%%%%%%%%%%%%%%%%%%
\subsection{LO state}%: $\Delta(x)= \Delta_0\cos(Qx)$}
\label{sec:LOsimple}

For the Larkin-Ovchinnikov (LO) state, the quadratic component in Eq.~(\ref{gen glf}) includes two terms identical to Eq.~\eqref{alpha FF bare} but summed over $\vq=\pm\vQ$ with $\Delta_{\pm\vQ} = \Delta_0/2$. Both terms for LO are identical when averaged over the Fermi surface, hence $\alpha_{LO}=\alpha_{FF}/2$. Thus the second order transition line, $B_{c}$, determined from $\alpha=0$, is identical for both the FF and LO phases.  The relative stability of the FF and LO phases is determined by comparing the quartic coefficients $\gamma_{FF}$ and $\gamma_{LO}$ at the transition, with the smaller of the two corresponding to the thermodynamically stable SC state because $f_{SC}-f_{N}=-\alpha^2/(2\gamma)$.

The quartic coefficient, $\gamma_{LO}$,  is obtained by summing the six terms in Eq.~(\ref{gamma general}) with $\vq_i\in\{\vQ,-\vQ\}$, subject to the constraint $\delta_{\vq_1+\vq_3,\vq_2+\vq_4}$. This yields
\begin{equation}\label{gamma LO bare}
\gamma_{LO}=t\re\hspace{-0.5mm}\left<\sum_{n=0}^\infty|\cY(\theta)|^4\frac{\bar{\omega
}_{n,b} \left(3 \bar{\omega
   }_{n,b}^2-\bar{q}^2\right)}{128 \pi ^2 \left(\bar{q}^2+\bar{\omega
   }_{n,b}^2\right)^3}\right>_{\hspace{-2mm}FS}\,,
\end{equation}
where $\bar{\omega}_{n,b} = t(n+\frac{1}{2})+i b$. Twenty distinct terms contribute to the sixth order Landau coefficient which becomes
\begin{equation}\label{nu LO bare}
\begin{split}
\nu_{LO}&=-t\re\hspace{-0.5mm}\left<\sum_{n=0}^\infty|\cY(\theta)|^6\right.\\
&\hspace{-1mm}\times\hspace{-1mm}\left.\frac{\bar{\omega
}_{n,b}\hspace{-1mm} \left(\bar{q}^6 -33 \bar{\omega }_{n,b}^2
   \bar{q}^4 +35 \bar{\omega }_{n,b}^4 \bar{q}^2 +5 \bar{\omega
   }_{n,b}^6\right)}{2048 \pi ^4 \left(\bar{q}^2+\bar{\omega
   }_{n,b}^2\right)^5 \left(9 \bar{q}^2+\bar{\omega
   }_{n,b}^2\right)}\right>_{\hspace{-2mm}FS}.
\end{split}
\end{equation}

We can obtain the gradient expansion of the free energy by expanding Eqs.~\eqref{alpha FF bare}-\eqref{nu LO bare} in powers of $q$ with the corresponding $\cY(\theta)$. The resulting Ginzburg-Landau expansion coefficients are identical to those obtained for $s$- and $d$-wave SC in Refs.~[\onlinecite{BuzdinFFLO:1997}] and [\onlinecite{SamokhinFFLO:1997}], respectively. Below, however, we retain the full $q$-dependence of the expansion coefficients to examine the transition line at low-$T$ where $Q \simeq \xi_0^{-1}$ and the gradient
expansions~\cite{BuzdinFFLO:1997,SamokhinFFLO:1997,BeairdFluctsGrad:2008}
fail.

%%%%%%%%%%%%%%%%%%%%%%%%%%%%%%%%%
\begin{figure}[t]
\includegraphics[width=0.45\textwidth,clip=true]{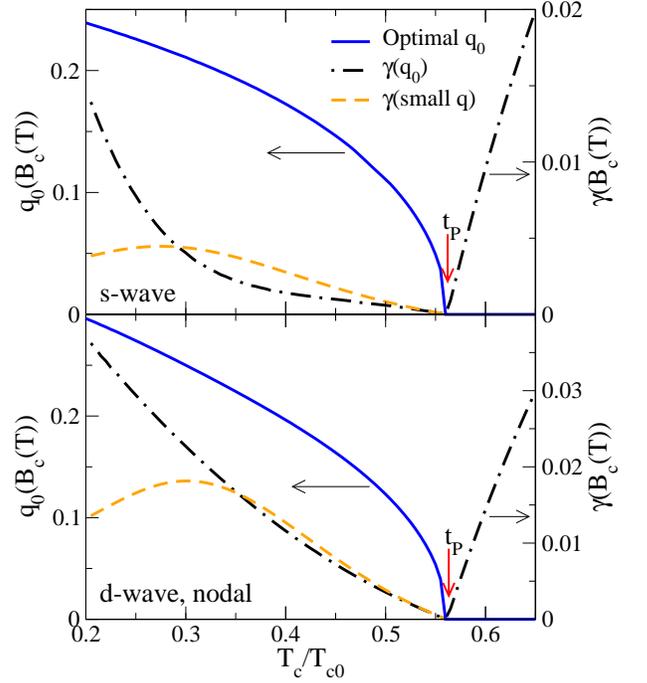}%
\vspace{-4mm}\caption{(Color online.) Optimal wave vector
$q_0(b_c(T))$ (solid) and quartic coefficient $\gamma(B_c(T))$
(dot-dashed) obtained by unrestricted maximization of $b_c(t)$ for LO state. Quartic coefficient for small-$q$ expansion (dashed) is also shown. Upper/lower panels are for $s$ and $d$-wave ($q$ along node) gaps. }\label{Fig:Q and Gamma plot}
\end{figure}
%%%%%%%%%%%%%%%%%%%%%%%%%%%%%%%%%%

\subsection{Determination of $B_c(T)$}

For each phase, with the free energy density written as
\begin{equation}\label{f gen form}
f =\alpha(t,b,q)|\delta_0|^2 +\gamma(t,b,q)|\delta_0|^4+\nu(t,b,q)|\delta_0|^6\,,
\end{equation}
we determine the critical field $B_c(T)$ and the optimal modulating wave vector $q_0$. We allow for possible second and first order transitions, and compare the results to determine the order of the physical transition.

The second order transition field at fixed $t$ is the maximal value of $b_c$ (with respect to $q$) for which $\alpha(t,b_c,q) = 0$, and $\gamma(t,b_c,q_0)>0$. The corresponding optimal $q_0$ determines whether the transition is into a uniform ($q_0=0$) or modulated state. In the vicinity of the transition line
\begin{equation}
|\delta_0|^2 =
-\frac{\alpha(t,b_c,q_0)}{2\gamma(t,b_c,q_0)}\approx\frac{\alpha^\prime
(t_c-t)}{2\gamma(t_c,b_c,q_0)}\,,
\end{equation}
where $\alpha^\prime=\partial\alpha(t,b_c,q_0)/\partial t|_{t=t_c}$. With this value we can compute the free energy difference between the normal and the superconducting states and therefore determine the thermodynamic properties such as the specific heat jump at the transition, see below.

In the region where $\gamma< 0$, the first order transition occurs once the minimum in the free energy shifts discontinuously to $\delta_0\neq 0$ before the quadratic coefficient $\alpha$ changes sign~\cite{ChaikinLubensky}.  This happens along the line defined by
\begin{equation}\label{first order criterion}
\gamma^2(t,b_c,q)-4\alpha(t,b_c,q)\nu(t,b_c,q)=0\,,
\end{equation}
where the new minimum first appears at
\begin{equation}\label{delta first order}
|\delta_0|^2 = -\frac{\gamma}{2\nu}\,.
\end{equation}

We locate the first order transition at a given $t$ by unrestricted maximization, with respect to $q$, of the field $b_c$ that satisfies Eq.~\eqref{first order criterion}.

At each temperature, we locate the maximal field $b_c(q)=b_c(q_0)$ for which the coefficient $\alpha(t,b_c,q_0)=0$. If we find $\gamma(t,b_c,q_0)>0$, the transition is second order. If $\gamma(t,b_c,q_0)<0$, we maximize $b_c$ for Eq.~\eqref{first order criterion}, checking that $\gamma$ remains negative and that the free energy remains bounded from below  ($\nu(t,b_c,q_0)>0$). The first and second order transition lines meet at a critical point $t^\star$ where $\gamma(t^\star,b_c,q_0)=0$. For $d$-wave gap, we compare the critical fields for nodal and antinodal orientations of the modulation wave vector $\vQ$ and verify that modulation along
gap nodes is preferred above $T\simeq0.06T_{c0}$.

As shown in Fig.~\ref{Fig:Q and Gamma plot} for LO modulation, the quartic coefficient remains positive and the transition is second order on both sides of $t_P$. For comparison, we also plot the results obtained from the small $q = \xi_0 Q/2$ expansion
\begin{equation}
\nonumber
f = (\alpha_0+\alpha_2 q^2 + \alpha_4 q^4)|\delta_0|^2\\
+(\gamma_0+\gamma_2 q^2)|\delta_0|^4 + \nu_0|\delta_0|^6\,
\end{equation}
of Eq.~\eqref{f gen form}, where the coefficients are found by expanding Eqs.\eqref{alpha general} and \eqref{gamma general}, and only even powers of $q$ appear since the system is isotropic. Since each subsequent term in the expansion contributes an extra $qG^0_\sigma$ in the Matsubara summation of Eqs.(\ref{alpha general})-(\ref{gamma general}), we have $\alpha_n\propto\gamma_{n-2}\propto \re(\Psi^{(n)}(\frac{1}{2}+i\frac{b}{t}))$ for $n\ge 2$. Consequently $\alpha_2$ and $\gamma_0$ change sign at exactly the same temperature, $t_P=T_P/T_{c0}$, and the modulated state with $q_0^2 = -\alpha_2/(2\alpha_4)$ emerges at lower $T$ via a second order transition described by
\begin{equation}\label{gradient expansion result}
f =
\left(\alpha_0-\frac{\alpha_2^2}{4\alpha_4}\right)|\delta_0|^2 +
\left(\gamma_0-\frac{\gamma_2\alpha_2}{2\alpha_4}\right)|\delta_0|^4
+ \nu_0|\delta_0|^6
\end{equation}
with the renormalized quartic term positive. The results obtained from an examination of Eq.~\eqref{gradient expansion result} are identical to those discussed in
Refs.~[\onlinecite{BuzdinFFLO:1997,SamokhinFFLO:1997}]. As is seen from Fig.~\ref{Fig:Q and Gamma plot}, below $t_P$ the transition is into modulated state with the wave vector that reaches values of $q_0\approx0.2\xi_0^{-1}$ and higher. The failure of the gradient expansion is manifested in the significant discrepancy between the values for the quartic coefficient at the optimal wave vector within the gradient expansion and in the full evaluation, shown in Fig.~\ref{Fig:Q and Gamma plot}.  Hereafter, we rely on unrestricted maximization of $b_c(t)$ with respect to $q$ to determine the critical field.

%%%%%%%%%%%%%      Magnetic Fluctuations     %%%%%%%%%%%%
\section{Magnetic fluctuations}

\subsection{Magnetic fluctuations}

If we are close to a magnetically ordered state, soft magnetic modes exist in the system. In the continuum limit the fluctuations of the magnetization field $M({\vr})$ are described by the Gaussian free energy
\begin{equation}
  \label{Eq:Fmag}
  \cF_{M}\left[\vM\right]=\frac{1}{2}\int
d\vr {\chi}^{-1} \vM(\vr)^2\,.
\end{equation}
Here we have ignored the the momentum dependence of $\chi({\vq})$, assuming the that the momenta relevant for superconductivity are $Q\sim \xi_0^{-1} \ll \pi/a$ (where $a$ is the lattice spacing). We do not discuss here the role of the (possibly singular) antiferromagnetic (AFM) fluctuations in mediating the superconducting pairing: this role can only be addressed within the framework of specific microscopic
theories\cite{MonthouxHiTcAFMCuOxides:1991,MoriyaAFMflucts2DHiTc:1990,NakamuraSpinFluctSC2d3d:1996}.
Our task is to consider the competition of superconductivity and the long-wavelength fluctuations of the magnetization, whether uniform or staggered. Although the susceptibility is enhanced ($\chi\gg\chi_P$), the system is not close to ferromagnetic order ($\chi\ll1$), hence we do not distinguish between $B$ and the applied magnetic field $H$ for the rest of the paper. In the same spirit, we ignore $\vM\cdot\vH$ in the magnetic free energy since its contribution to the averaged free energy is a factor of $\chi$ smaller than the corrections we consider.

In Ref.~[\onlinecite{KMV:2003}] the susceptibility was taken to be temperature-dependent, in agreement with experiment~\cite{Tayama:2002}, $\chi(T) = \chi_0 T_{sf}/(T+T_{sf})$ where $T_{sf}$ is a characteristic energy scale for the low-energy spin fluctuations. While we make use of this expression to make contact with Ref.~[\onlinecite{KMV:2003}], our main results are
qualitatively the same for a temperature independent $\chi$ of the same magnitude. We also ignore the field dependence of $\chi$. Finally, we do not account for the quantum fluctuations of $\vM$ and consider only thermal fluctuations of the magnetization.

Recall that our goal is to investigate the effects of long-wavelength magnetic fluctuations on the N-SC transition. We include the competition between magnetism and superconducting orders via the lowest order term allowed by symmetry in the free energy expansion,
\begin{equation}\label{Fsc,m}
\cF_{sc,M}\left[\Delta,\vM\right]=\frac{\eta}{2}\int d\vr
|\Delta(\vr)|^2 \vM(\vr)^2\,,
\end{equation}
where the coupling constant $\eta>0$ makes coexistence of the two orders unfavorable. In a simple system it would be possible to determine $\eta$ from microscopics, by expanding ${\vB}={\vH}+4\pi{\vM}$ in each Green's function in the powers of the fluctuating magnetization and introducing the correlator $\langle\vM(\vr)\vM(\vr^\prime)\rangle$
that is proportional to the susceptibility, in analogy with
Ref.~[\onlinecite{LeeFluctsPierlsTrans:1973}]. Such an expansion produces an $M^2|\Delta|^2$ term. In a complex system with $f$-electrons we cannot determine the coefficient of this term microscopically, and we use Eq.~\eqref{Fsc,m} with a phenomenological parameter $\eta$ to explore the salient features of the model. Our consistency checks on the choice of $\eta$ are the magnitudes of the jump in $\Delta_0$ and of $\Delta M/M$ across the first order transition line. We find maximal $\Delta_0(T_c)\lesssim0.3\Delta(0)$ and $\Delta M/M\approx1-5\%$ everywhere along the first order transition line. These values are moderate, hence our choice of $\eta$ is physically reasonable.

To verify the ubiquity of our results, we examined the coupling of the SC order parameter to higher order terms in $M(\vr)^2$ and its gradients, \emph{e.g.}
$|\Delta(\vr)|^2 |\nabla M(\vr)|^2$, and, within the small-$q$ approximation, to gradients of the order parameter itself, \emph{e.g.} $|\nabla \Delta(\vr) \cdot \vM(\vr)|^2$.  We checked that, while these various couplings renormalize the transition temperature, they do not introduce new features into the phase diagram.

To derive the effective theory for the superconducting order we integrate out the magnetic fluctuations from the partition function
\begin{equation}
\nonumber \cZ=\exp\left[-(\cF_{L}+\cF_{sc,M} +\cF_M)/T\right]\equiv
e^{-\cF_{L}/T}\overline\cZ_{sc,M}\,
\end{equation}
where $\cF_{L}=\int d^D\vr F_{L}$. We obtain the total free energy
\begin{equation}
\overline{\cF} =  \cF_{L}-T\ln{\overline\cZ_{sc,M}}\,,
\end{equation}
where the magnetic contribution is
\begin{equation}\label{Zsc,m}
\overline\cZ_{sc,M}=\int\cD[\vM(\vr)]\exp\left[-\frac{1}{T}
\left(\cF_{sc,M}+ \cF_{M}\right)\right]\, ,
\end{equation}
and ${\cal D}[\vM(\vr)]$ indicates integration over all possible configurations of magnetization.  The integral is Gaussian in $\vM$, hence we compute it analytically and expand in powers of $|\Delta|^2$ to obtain the corrections due to magnetic fluctuations to the expansion coefficients in $F_{L}$. Below we address these corrections in each of the three phases we consider: uniform, FF, and LO.

%%%%%%%%%%%%%%%%%%%%%%%%%%%%%%%%%%%%%%%%%%%%%%%%%%%%%%%%%%%%%%%%%%%%%%%%%%%%%%%%%%%%%%%%%
\subsection{Uniform superconducting state}

Integrating out the fluctuating magnetization for a uniform order parameter, $\Delta_0$, is straightforward. We work with the Fourier components of the magnetization, $\vM_\vk$, and restrict the sum
\begin{equation}\label{uni k-space energy}
\cF_{sc,M}+ \cF_{M}= \sum_{|\vk|<k_c} \frac{1}{2}\left(\chi^{-1}+
\eta |\Delta_0|^2\right) |\vM_\vk|^2\,,
\end{equation}
to one-half of $k$-space since $\vM_\vk=\vM_{-\vk}^\star$ for real $\vM(\vr)$. Therefore from Eq.~\eqref{Zsc,m} we have, after Gaussian integration over both real and imaginary parts of $\vM_\vk$,
\begin{equation}\label{Zuni integrated}
\overline\cZ_{sc,M}
=\prod_{|\vk|<k_c}\left(\frac{2\pi\chi
T}{1+\eta\chi|\Delta_0|^2}\right)^\frac{d}{2}
\end{equation}
where now the product is taken over all $\vk$ up to the cutoff of the order of the lattice spacing $|\vk_c|=\pi/ l$, and $d$ is the dimensionality of magnetization vector $\vM$.

Neutron scattering\cite{BroholmSpinResCeCoIn5:2008} measurement of the dynamic spin susceptibility in CeCoIn$_5$ shows evidence of spin fluctuations, and light Cd-doping\cite{PhamCdDoping:2006} induces AFM order at $Q_{AFM}=(.5,.5,.5)$\cite{NicklasCdDopedCeCoIn5:2007}. Sister compound CeRhIn$_5$ exhibits AFM order at $Q_{AFM}=(.5,.5,.297)$\cite{BaoIncomAfmCeRhIn5:2000}, which is stable under pressure\cite{MajumdarPressAfmCeRhIn5:2002, LlobetMagStructCeRhIn5Pressure:2004}, before SC preempts AFM order at $P\approx2$GPa\cite{HeggerPressScCeRhIn5:2000}. Furthermore, the pressure dependence of $T_c$ and $T_N$ for CeRhIn$_5$ and Cd-doped CeCoIn$_5$ is nearly identical\cite{PhamCdDoping:2006} suggesting that the SC and magnetic orders in both are closely related. Hence we conclude that CeCoIn$_5$ is in proximity to 3D magnetic ordering, and we take $d=3$ for the purposes of this paper.

The corresponding contribution to the free energy is $\cF_M(\Delta_0)=-T\ln\overline\cZ_{sc,M}$. Subtracting the average magnetic contribution to the normal state energy, $\overline{\cF}_{M}(\Delta_0=0)$ we find an additive contribution to the superconducting free energy density
\begin{equation}\label{Funi flucts}
\begin{split}
F_{uni,M} &= \frac{\overline{\cF}_{uni,M}(\Delta_0)-\overline{\cF}_{uni,M}(0)}{L^D}\\
&=\frac{3}{2}\frac{T}{L^D}\sum_{k<k_c}\ln\left(1+\eta\chi|\Delta_0|^2\right)\\
&=\frac{3}{2}\frac{T}{l^D}\ln\left(1+\eta\chi|\Delta_0|^2\right)
\end{split}
\end{equation}
where $L^D$ is the volume. The last line of Eq.~\eqref{Funi flucts} follows from $\sum_k a\approx a(L/l)^D$ where $a$ does not depend on $k$.  Expanding this contribution in powers of $|\Delta_0|$ for our 2D superconductor ($D=2$), we find the renormalized coefficients of $f=F/(N_FT_{c0}^2)$, Eq.~(\ref{f gen form}),
\begin{subequations}\label{eq:GLu}
\begin{eqnarray}
\label{alpha u}
\overline{\alpha}_u=\alpha_u+\frac{3}{2}t\frac{T_{c0}}{N_F l^2}\eta\chi(T), \\
\label{gamma u}
\overline{\gamma}_u=\gamma_u-\frac{3}{4}t\frac{T_{c0}^3}{N_F l^2}\eta^2\chi^2(T), \\
\label{nu u} \overline{\nu}_u=\nu_u+\frac{1}{2}t\frac{T_{c0}^5}{N_F
l^2}\eta^3\chi^3(T)\,,
\end{eqnarray}
\end{subequations}
where $\alpha_u,\gamma_u$, and $\nu_u$ are given in Eqs.~(\ref{alpha u bare})-(\ref{nu u bare}).

Since in the FF state, $\Delta(x) = \Delta_0 e^{i Q x}$, only the phase of the order parameter is modulated, the coupling between the magnetization and the superconducting order, Eq.~(\ref{Fsc,m}), has exactly the same form as in the uniform state. Hence the renormalized expansion coefficients are obtained from Eqs.~\eqref{eq:GLu} by a direct substitution of
$\alpha_{FF}$, $\gamma_{FF}$, and $\nu_{FF}$ for $\alpha_u$,
$\gamma_u$, and $\nu_u$, respectively.

%%%%%%%%%%%%%%%%%%%%%%%%%%%%%%%%%%%%%%%%%%%%%%%%%%%%%%%%%%%%%%%%%%%%%%%%%%%%%
\subsection{Modulated LO state}

In the LO state, in addition to the order parameter
$\Delta(x) = \Delta_0\cos(Qx)$ competing with the average magnetization, the amplitude modulation couples the magnetic fluctuations at wave vectors differing by $2Q$. Therefore, the magnetic contribution to the free energy is
\begin{equation}\label{mod k-space energy}
\begin{split}
\cF_{sc,M}+ \cF_{M}&= \sum_{|\vk|<k_c}
\left[\frac{1}{2}\left({\chi}^{-1}+\frac{1}{2}\eta|\Delta_0|^2\right)
|\vM_\vk|^2\right.\\
&\left.-\frac{1}{8}\eta |\Delta_0|^2\vM_\vk
\cdot\left(\vM^\star_{\vk+2\vQ}+\vM^\star_{\vk-2\vQ}\right)\right]\,.
\end{split}
\end{equation}
After integrating out the fluctuations, the contribution to the superconducting free energy density relative to the normal state becomes
\begin{equation}\label{Fmod flucts}
\begin{split}
\overline F_{LO,M}
&=\frac{3}{2}\frac{T}{L^D}\sum_{k<k_c}\left[\ln\left(1+\frac{1}{2}\eta\chi|\Delta_0|^2\right)\right.\\
&+\left.\ln\left(1-\frac{1}{8}\eta^2\chi^2|\Delta_0|^4+\frac{1}{8}\eta^3\chi^3|\Delta_0|^6\right)\right].
\end{split}
\end{equation}
The first term differs from its counterpart in
Eq.~\eqref{Funi flucts} by the factor of 1/2, arising from the spatial average of $\cos^2(Qx)$. The second term arises from the mode-mode coupling terms in Eq.\eqref{mod k-space energy} and is derived in Appendix~\ref{app:TriDiagInt}. Under expansion in $\Delta_0$, it only contributes to the fourth and sixth order terms in the free energy, and we obtain
\begin{subequations}\label{eq:GLlo}
\begin{eqnarray}
\label{alpha LO}
\overline{\alpha}_{LO}=\alpha_{LO}+\frac{3}{4}t\frac{T_{c0}}{N_F l^2}\eta\chi(T), \\
\label{gamma LO}
\overline{\gamma}_{LO}=\gamma_{LO}-\frac{3}{8}t\frac{T_{c0}^3}{N_F l^2}\eta^2\chi^2(T), \\
\label{nu LO}
\overline{\nu}_{LO}=\nu_{LO}+\frac{1}{4}t\frac{T_{c0}^5}{N_F
l^2}\eta^3\chi^3(T)\,,
%\end{equation}
\end{eqnarray}
\end{subequations}
where $\alpha_{LO},\gamma_{LO}$, and $\nu_{LO}$ are given in Sec.~\ref{sec:LOsimple}.

Comparing Eqs.~\eqref{eq:GLu} and \eqref{eq:GLlo}, we see that the free energy expansion depends on $\eta$ and $\chi$ only through their product $\eta\chi$. Thus, for subsequent analysis we define a dimensionless coupling parameter
\begin{equation}\label{tEta}
\tEta = \frac{3}{2}\frac{T_{c0}}{N_Fl^2}\eta = \frac{3}{2}T_{c0}T_F\eta
\end{equation}
where the characteristic temperature $T_F=(N_Fl^2)^{-1}$ is of the order of the Fermi temperature in the system. We also define a dimensionless parameter based on the experimental fit of $\chi(T)=\chi_0 T_{sf}/(T+T_{sf})$
\begin{equation}
\tChi =\chi_0 \frac{T_{sf}}{T_{c0}}\, .
\end{equation}
With these parameters, the renormalized quadratic coefficients in Eqs.~\eqref{alpha u} and \eqref{alpha LO} become simpler,
\emph{e.g.},
\begin{equation}\label{alpha with dimless params}
\bar{\alpha}_u= \alpha_u+ \tEta\tChi\frac{t}{t + t_{sf}} \,.
\end{equation}
The renormalization of all other Landau coefficients is determined by the product $\tEta\tChi$, and, in simplifying the fourth and sixth order terms in Eqs.~\eqref{eq:GLu} and \eqref{eq:GLlo}, we introduce the parameter $t_F = T_F/T_{c0}$.

We note that the dimensionality of the magnetization vector $\vM$ enters the Landau coefficients as a prefactor of the coupling parameter $\eta$. Throughout this paper, we take $d=3$. Using a different value for $d$ simply decreases the magnetic fluctuation contributions in Eqs.~\eqref{eq:GLu} and \eqref{eq:GLlo} by a factor of $d/3$. For example, taking $d=2$ only requires that we use $3\eta/2$ to obtain the same results (\emph{e.g.}, $T_c$) as for $\eta$ and $d=3$. Hence, we proceed with our choice $d=3$ without any loss of generality.

%%%%%%%%%%%%%%%%%%%%%%%%%%%%%%%%%%%%%%%%%%%%%%%%%%%%%%%%%%%%%%%%%%
\subsection{Choice of energy scales and parameters}
\label{sec:energy scales}

The exchange of entropy between the magnetic fluctuations  and superconductivity reduce the zero-field transition temperature from the unrenormalized $T_{c0}$ to the experimentally observed $T_c(\eta>0)$ as determined from the instability condition
\begin{equation}\label{zero B instability}
0 = \overline{\alpha}_u = \ln\left(\frac{T_c}{T_{c0}}\right)
+\frac{3}{2}T_cT_F\eta\chi\,.
\end{equation}
The extra entropy is released in a specific heat jump that exceeds the BCS value,
\begin{equation}
\frac{\Delta C/T_c(\eta)}{N_F T_{c0}^2}=\left.-\frac{\partial^2
f}{\partial T^2}\right|_{T_c(\eta)}
=\left.\frac{[\bar{\alpha}^\prime(\eta)]^2}{2\bar{\gamma}(\eta)}\right|_{T_c(\eta)}
\,,
\end{equation}
where $f(\eta)$ is the dimensionless free energy for the given coupling, $\eta$, and $\bar{\alpha}'=\partial\bar{\alpha}(T)/\partial T$. Without magnetic fluctuations, BCS mean field theory predicts for $s$-wave gap $\Delta C/C_N=12/7\zeta(3)\approx1.43$ and for $d$-wave gap $\Delta C/C_N=8/7\zeta(3)\approx0.95$ at $T_{c0}$.  Here $C_N$ is the normal state specific heat, and $\zeta(3)\approx 1.202$ is the Riemann zeta function. Measuring the jump relative to the $s$-wave value, we find for $B=0$
\begin{equation}\label{spec heat jump zero field}
\frac{\Delta C/T_c(\eta)}{1.43C_N/T_{c0}}
=\left.\frac{\left(1+\frac{3}{2}T_c T_F
\eta\left(\chi+T_c\chi^\prime\right)\right)^2}
{\left<|\cY(\theta)|^4\right>-\frac{3(2\pi)^2}{7\zeta(3)}T_c^3T_F\eta^2\chi^2}\right|_{T_c}
\end{equation}
where $T_c=T_c(\eta)$ and $\chi^\prime=\partial\chi(T)/\partial T$. Using Eq.~(\ref{zero B instability}) to eliminate $\eta$ we find
\begin{equation}\label{spec heat jump figure out}
\frac{\Delta C/T_c}{1.43C_N/T_{c0}}\hspace{-1mm}
=\frac{\left(1+\frac{\left(\chi+T_c\chi^\prime\right)}{\chi}\ln\left(\frac{T_{c0}}{T_c}\right)\right)^2}
{\left<|\cY(\theta)|^4\right>-\frac{4(2\pi)^2}{21\zeta(3)}\frac{T_c}{T_F}\ln^2\left(\frac{T_{c0}}{T_c}\right)}\,,
\end{equation}
in zero field. We discuss the field dependence of $\Delta C/T_c$ in Section~\ref{sec: spec heat jump}.

From the experimentally measured behavior of the susceptibility, specific heat jump $\Delta C/T_c(B=0)$, and $T_c$ one can estimate $T_{c0}$ provided a reasonable guess about the value of $T_F$ can be made. For our purposes, we take $T_{c}=2.3K$, $T_F = 40K$ (the Kondo coherence temperature for $\cecoin$\cite{NakatsujiKondoTemps:2002}), and the dimensionless $\chi_0\approx 10^{-4}$ (presented in units of $emu/g$ in Ref.~[\onlinecite{Tayama:2002}]). We follow the example of Ref.~[\onlinecite{KMV:2003}] and set $T_{sf}= 1.5K$. With this choice $T_{sf}<T_{c}$, and we examine the effects of $\chi$ which varies substantially with temperature below $T_c$. Experiment, however, suggests a weaker temperature dependence of $\chi(T)$ with $T_{sf}\approx 3.5T_{c}$\cite{Tayama:2002}. Therefore, we verify that our general results are independent of the details of $\chi$ by comparing this case with the analysis for constant susceptibility. For our chosen energy scales, we solve Eq.~\eqref{spec heat jump figure out} with $\Delta C/T_c=3\Delta C/T_{c0}$.  This gives for $s$-wave $T_{c0}=6.20K$ and $\tEta\tChi\simeq1.6$ and for $d$-wave $T_{c0}=9.27K$ and $\tEta\tChi\simeq2.3$.

%%%%%%%%%%%%%%%%%%%%%%%%%%%%%%%%%%%%%%%%%%%%%%%%%%%%%%%%%%%%%%%%%%%%%%%%%%%%%%%%%%%%%%%%
\section{Discussion and Results}

Using the formalism outlined above, we are now in the position to investigate the changes appearing in the transition lines of the superconductor coupled to the magnetic fluctuations. In the following we set $T_F$, $T_{sf}$, and $T_{c0}$ as described at the end of the previous section. We adjust the coupling $\eta$ to the magnetic fluctuations as well as the temperature dependence of the magnetic susceptibility. We first address the nature of the transition along the $B_c(T)$ line, and then consider the thermodynamic signatures of these transitions.

%%%%%%%%%%%%%%%%%%%%%%%%%%%%%%%%%%%%%%%%%%%%%%%%%%%%%%
\subsection{Normal to superconducting transition in a magnetic field. }

Quite generally coupling to magnetic fluctuations suppresses the transition temperature, since, as is clear from  Eq.~(\ref{Fsc,m}), the finite thermal average of $\vM^2(\vr)$ makes the appearance of superconductivity energetically costly. This is also evident from Eqs.(\ref{alpha LO}) and (\ref{alpha u}), which show positive additive contribution to the quadratic coefficients in the Landau expansion. In the absence of the  field, when $\alpha(T)=-\ln T/T_{c0}$, it follows from  Eq.(\ref{zero B instability}) that the transition temperature $T_c$ satisfies
\begin{equation}
\left.\frac{T_c}{T_{c0}}\right|_{B=0}=
e^{-\frac{3}{2}T_cT_F\eta \chi(T_c)}=
\exp\left(-\frac{\tEta\tChi\,T_c}{T_c+T_{sf}}\right)\,,
\end{equation}
where in the last step we explicitly invoked the temperature
dependence of the susceptibility. For small $\tEta\tChi$ the
linearized form of this equation coincides with that used in
Ref.~[\onlinecite{KMV:2003}].

%%%%%%%%%,clip=true, viewport=1.5cm 1.5cm 20cm 14.1cm %%%%%%%%
\begin{figure}[t]
\includegraphics[width=0.48\textwidth,clip=true, viewport=0.0cm 0.0cm 14cm 21.5cm]{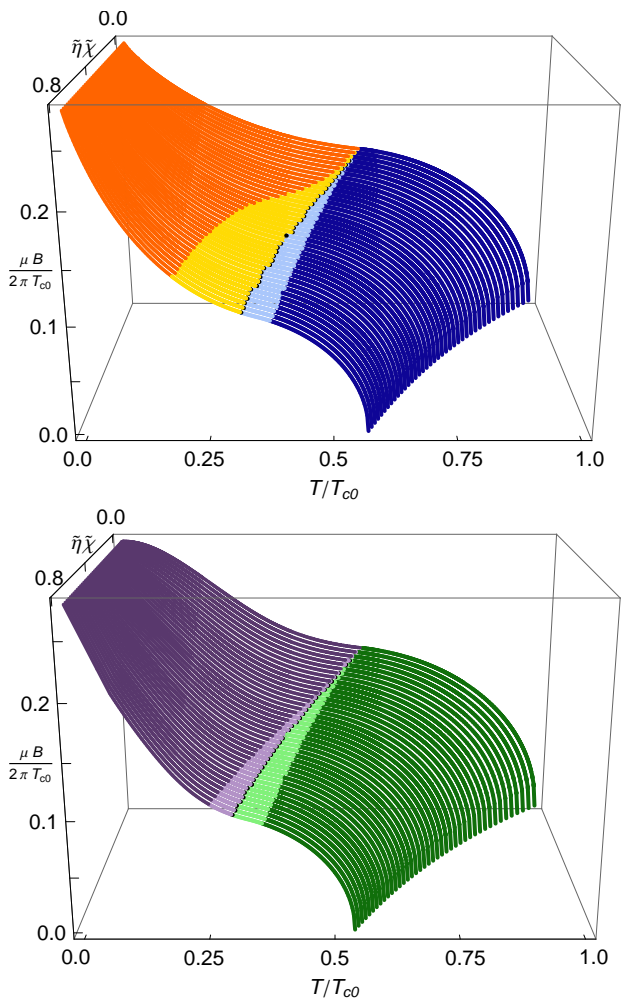}%
\caption{(Color online.) The normal-to-superconducting transition in $s$-wave (upper) and $d$-wave (lower) superconductors under a Zeeman field. Modulation is along gap node for $d$-wave. Magnetic fluctuations, $\tEta$, modify the $B_{c2}$ transition, showing 4 distinct regions (for increasing $T$): 2nd order into modulated, 1st order into modulated, 1st order into uniform and 2nd order into uniform states. \label{fig:3D diagram} }
\end{figure}
%%%%%%%%%%%%%%%%%%%%%%%%%%%%%

At the same time the results for the quartic coefficient,
Eqs.(\ref{gamma LO}) and (\ref{gamma u}) show that it is
renormalized {\em downward} by the magnetic fluctuations. Since the sign of this term controls whether the transition is of the second or first order, it seems possible that the order of the
transition may change as the strength of the magnetic fluctuations increases.

%%%%%%%%%%%%%%%%%%%%%%%%%%%%%
\begin{figure}[t]
\includegraphics[width=0.45\textwidth,clip=true]{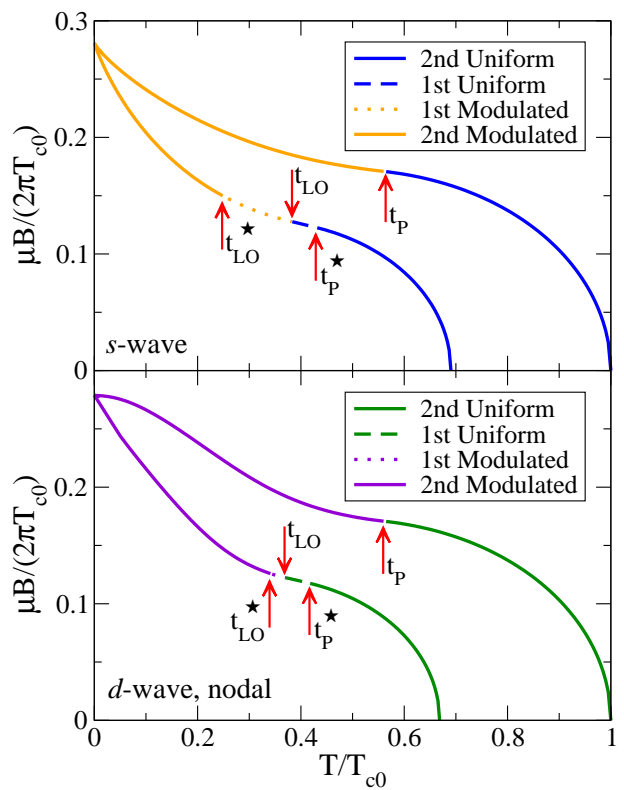}%
\caption{(Color online.) Transition lines for $s$- and $d$-wave with nodally-oriented $\vQ$. Upper curves are in the absence of fluctuations, and lower curves are for $\tEta\tChi=0.5$.
\label{fig:phase diagram compare} }
\end{figure}
%%%%%%%%%%%%%%%%%%%%%%%%%%%%%

Figures~\ref{fig:3D diagram} and \ref{fig:phase diagram compare} show that this is indeed the case: coupling to magnetic fluctuations opens a region of first order transition from the normal to both uniform and the modulated superconducting state. This finding is a major conclusion of our work, and qualitatively fits with the behavior of CeCoIn$_5$ where the transition becomes first order below $T_1\approx 1$K\cite{BianchiFOTCeCoIn:2002}, while the putative FFLO-like phase does not occur until a lower temperature\cite{BianchiFFLO:2003}.

To understand this behavior recall that in the
absence of fluctuations~\cite{Maki:1964, BuzdinFFLO:1997, SamokhinFFLO:1997} the quartic term of the Ginzburg-Landau expansion for the uniform superconducting phase changes sign,
$\gamma_u (T_P)=0$ exactly at the point along the $B_c(T)$ line (at temperature $T_P$) where the modulated phase, reached via a second order transition, $\alpha_{LO}(T_P)=0$, becomes allowed. Coupling to the fluctuations increases $\alpha_{LO}$ and lowers $\gamma_u$ ensuring that the first order transition in the uniform state occurs at higher temperature than that where the modulated phase can form.

As is seen from Figs.~\ref{fig:3D diagram} and \ref{fig:phase
diagram compare} the region of the first order transitions widens as the fluctuations become softer ($\chi$ increases) or compete more strongly ($\eta$ increases) with superconductivity. There we define the temperatures $t_P^\star$ and $t_{LO}^\star$ where the second order transitions into the uniform and LO modulated superconducting states respectively become first order.

Since we use the expansion in powers of $\delta_0$ we can only estimate the location of the first order transition line away from the critical points at which the transition becomes second order. However, since the jump in $\delta_0$ across the first order transition is modest (\emph{e.g.}, for $s$-wave gap, $\delta_0(t_c)\lesssim0.3\delta_0(0)$ with $\delta_0(0)=\pi e^{-\gamma_E}\approx1.76$) this estimate is quite reliable. We denote by $t_{LO}$ our estimate of the temperature along the $b_c(t)$ where the first order transition lines into the uniform and the LO phases meet. For $t<t_{LO}$ the transition (first or second order) is into the amplitude-modulated phase, while for $t>t_{LO}$ it is into a uniform phase.  In the absence of fluctuations, of course,
$t_P^\star=t_{LO}^\star=t_{LO}=t_P$.

%%%%%%%%%%%%%%%%%%%%%%%%%%%%%%%%%
\begin{figure}[t]
\includegraphics[width=0.45\textwidth,clip=true]{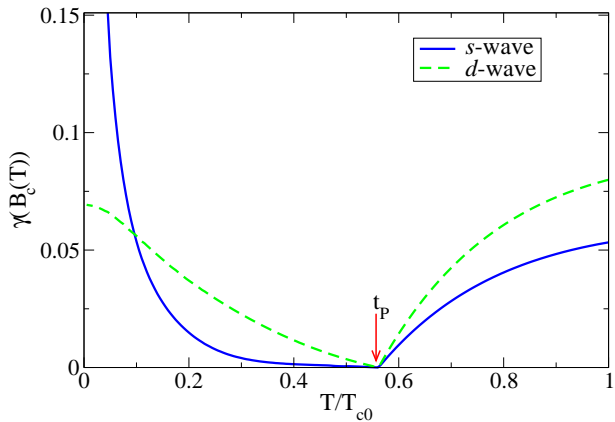}%
%\vspace{4mm}
\caption{(Color online.) Quartic Landau coefficient evaluated along $B_c(T)$ for $s$-wave and $d$-wave symmetries with LO modulation for $t< t_P$ and uniform state for $t>t_P$.} \label{fig:Gamma Usual}
\end{figure}
%%%%%%%%%%%%%%%%%%%%%%%%%%%%%%%%%%

We find that for $s$-wave order the region of the first order
transition, for the same values of the coupling and magnetic
susceptibility, is wider. This can be qualitatively explained by examining the quartic Landau coefficient for both symmetries in the absence of fluctuations, shown in Fig.~\ref{fig:Gamma Usual}. In the vicinity of $T_P$, the coefficient $\tilde{\gamma}$ is numerically smaller for an $s$-wave order parameter than for $d$-wave, both on the uniform and the modulated (with the wave vector $Q$ yielding maximal $B_c$ for each symmetry) side of the transition. Hence it is easier to drive an $s$-wave system to first order transition.

Note that for  $d$-wave SC we find that the modulation of the
order parameter along the gap nodes is stabilized even below
$T=0.06T_{c0}$, where, in the absence of fluctuations, the
anti-nodal direction would be more
advantageous~\cite{MakiSineLikeDwave:1996,YangSondhi:1998,AntonFFLO2D:2005}.
The anti-nodal modulation still gives a lower free energy at very low temperatures, below a threshold that depends on the parameter $\tEta\tChi$, but that occurs far from the first order transition range that is our focus here, and therefore for the rest of this paper, we discuss only $d$-wave SC where $\vQ$ is oriented along a gap nodes.

\begin{figure}[t]
\includegraphics[width=0.45\textwidth,clip=true]{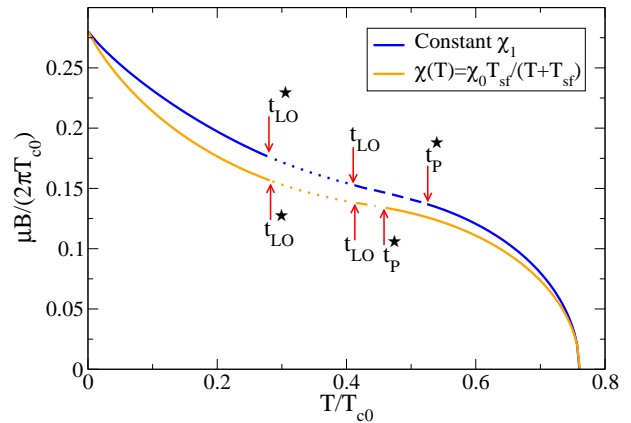}%
%\vspace{-4mm}
\caption{(Color online.) $S$-wave $b_c(t)$ for both constant $\chi(t)$ and $\chi_1\equiv\chi(t_c)$ for $t_c=0.76$. Upper and lower curves are for $\chi_1$ and $\chi(t)$, respectively. The critical field is suppressed less for $\chi_1$ since $\tEta\chi_1<\tEta\chi(t)$ at $t<t_c$. As $\tEta\chi<1$ for both cases, the region of first order transitions is larger for $\chi_1$.}
\label{fig:Chi Comp Plots}
\end{figure}

The key finding of the region of the first order transition does not depend on the exact temperature dependence of $\chi(T)$. For comparison, we also considered the constant susceptibility $\chi_1 \equiv \chi(T_c)$ so that, for a given coupling strength $\tEta$, we obtain the same $T_c$. In Fig.~\ref{fig:Chi Comp Plots}, we compare the critical field and order of transition for $\chi(T)$ and constant $\chi_1$. Since $\tEta\chi_1<\tEta\chi(T)$ for all $T<T_c$, superconductivity is suppressed \emph{less} and $B_c(T)$ is higher for constant susceptibility. However, in both cases the product $\tEta\chi<1$, and the magnetic fluctuations have a \emph{larger} effect for constant susceptibility than for $\chi(T)$ on the fourth Landau coefficient where $\tEta\chi$ enters quadratically. Thus, the N-SC transition is first order over a wider temperature range for constant susceptibility. While both the exact temperature range of first order transition and the degree of $B_c(T)$ suppression depends on the temperature dependence of $\chi$, the \emph{presence} of these effects is independent of the details of the susceptibility. Furthermore, the thermodynamics of the transition are similar for both $\chi_1$ and $\chi(T)$ where the only significant difference is the low-$T$ behavior of the specific heat jump for $d$-wave as discussed below.

\begin{figure}[t]
\includegraphics[width=0.45\textwidth,clip=true]{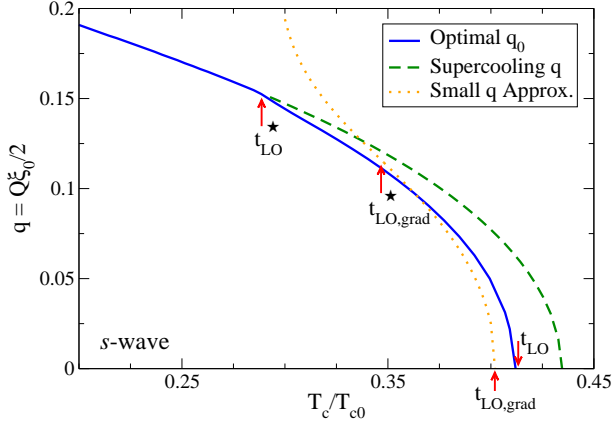}%
%\vspace{-4mm}
\caption{(Color online.) Optimal wave-vector at N-SC transition for $s$-wave with $\tEta\tChi=0.4$. Small-$q$ approximation predicts $t_{LO,grad}$ and $t^\star_{LO,grad}$. Upswing in $q$ below $t^\star_{LO,grad}$ indicates breakdown of small-$q$ approximation\cite{BeairdFluctsGrad:2008}.}
\label{Fig:Q0compare}
\end{figure}

We note that our results agree with those obtained from a small-$q$ expansion of the free energy functional\cite{BeairdFluctsGrad:2008}. In Fig.~\ref{Fig:Q0compare} we compare the results obtained from the fully $q$-dependent Landau functional and the small-$q$ approximation by plotting the optimal wave vectors found via each method. Each model predicts first and second order transitions into both the uniform and modulated SC states. Hence, our current model supports our preliminary results\cite{BeairdFluctsGrad:2008} while allowing us to examine the upper critical field beyond the limitations of a small-$q$  approximation.

%%%%%%%%%%%%%%%%%%%%%%%%%%%%%%%%%%%%%
%%%%%%%%%%%%%%%%%%%%%%%%%%%%%%%%%%%%%

Our main conclusion so far is therefore that coupling to thermal magnetic fluctuations drives the transition from the normal to superconducting state first order in the vicinity of the onset of the modulated state. Importantly, the transition is first order on both sides of this point, i.e. we find first order transitions both in the uniform and into the LO state. At lower temperatures the transition to the inhomogeneous superconducting state is second order. This is natural within our picture since the thermal fluctuations ``die out'' as the temperature is lowered. Within the present framework we cannot determine whether, should the quantum dynamics of the magnetization be accounted for, the transition would remain first order to the lowest temperatures. However, since $t_{LO}^\star\approx0.5t_c$ for $d$-wave order parameter (Figs.~\ref{fig:3D diagram} and \ref{fig:phase diagram compare}), it appears likely that the LO transition becomes second order again at high enough temperatures so that the quantum fluctuations are unlikely to have a major effect. We now investigate the thermodynamic signatures of these transitions.

%%%%%%%%%%%%%%%%%%%%%%%%%%%%%%%%%%%%%%%%%%%%%%%%%
\subsection{Thermodynamics at N-SC transition}

\subsubsection{Specific heat jump at the second order transition.}\label{sec: spec heat jump}

%%%%%%%%%%%%%%%%%%%%%%%%%%%%%%%%%
%%%%%%%%%%%%%%%%%%%%%%%%%%%%%%%%%%

\begin{figure}[t]
\includegraphics[width=0.45\textwidth,clip=true]{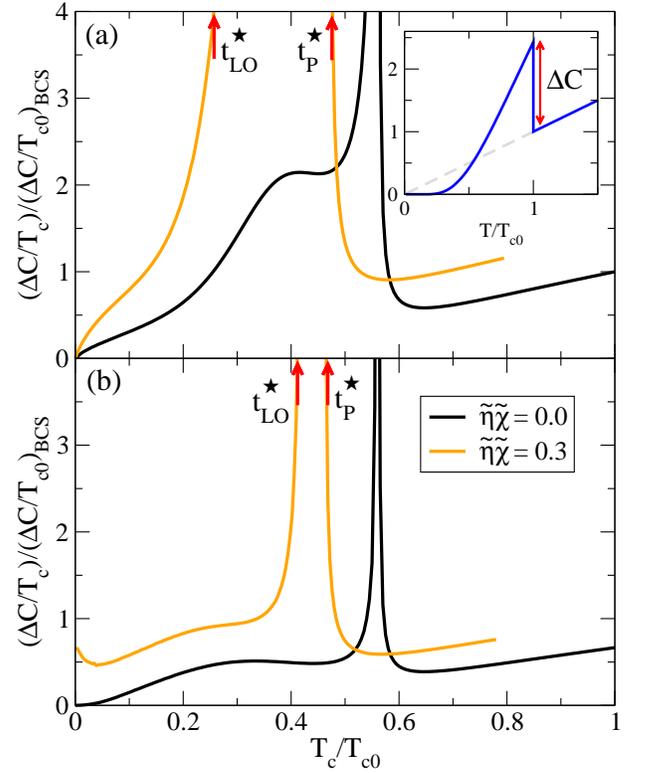}%
%\vspace{-4mm}
\caption{(Color online.) Specific heat jump at second order N-SC transition for (a) $s$-wave and (b) $d$-wave symmetries with $\tEta\tChi=$0.0, 0.3. Inset: Specific heat jump $\Delta C$ at second order N-SC transition.}
\label{fig:Spec Heat}
\end{figure}

The specific heat jump, measured relative to the BCS $s$-wave value, at the second order N-SC transition along $B_c(T)$ is given by
\begin{equation}\label{spec heat wrt BCS}
\frac{\Delta C/T_c(\eta)}{1.43 C/T_{c0}} =
\frac{7\zeta(3)}{8\pi^2}
\left.\frac{(\bar{\alpha}^\prime)^2}{2\bar{\gamma}}\right|_{T_c,B_c,Q_0}\,.
\end{equation}
Here again the prime denotes the temperature derivative, and the quadratic and quartic coefficients are determined from Eqs.~\eqref{eq:GLu} and \eqref{eq:GLlo} evaluated at the transition point and optimal modulation vector $Q_0$. The results are presented in Fig.~\ref{fig:Spec Heat} for the $s$ and $d$-wave superconductors.

Not surprisingly, the specific heat jump diverges on approaching the first order transition region. Note that in the absence of fluctuations, even though the transition remains second order throughout, there is a singularity in $\Delta C/T_c$ due to the vanishing of the quartic coefficient at $T_P$. The shoulder in the specific heat in the modulated state is found both with and without coupling to the magnetic moment, and hence simply reflects the details of the variation of the coefficients and the modulation wave vector with temperature.

Of more interest is the low temperature behavior. While for
$s$-wave superconductors the specific heat jump vanishes as
$T\rightarrow 0$ for both $\eta=0$ and $\eta\neq 0$, for the
$d$-wave symmetry the same jump is a) finite for $\eta\neq 0$, and b) exhibits a minimum at the lowest $T$.

The key to understanding this behavior is in evaluating the $T=0$ limit of the coefficients $\alpha_{LO}$ and $\gamma_{LO}$, which can be done analytically as detailed in Appendix~\ref{app:Zero T limit}. Note that the classical fluctuations disappear at $T=0$, as evidenced by the linear in $t$ fluctuation corrections in Eqs.~\eqref{eq:GLu} and \eqref{eq:GLlo} and that the values of  $b_c$ and $Q_0$ at $t=0$  do not depend on $\eta$ or $\chi$. For $s$-wave symmetry, the optimal wave vector and critical field are $Q_{0,s}=e^{-\gamma_E}\xi_0^{-1}\approx
0.56\xi_0^{-1}$ and $b_{c,s}=e^{-\gamma_E}/2\approx0.28$ ($\gamma_E\approx0.577$ is Euler's constant) at zero temperature, respectively. We find that for the $s$-wave case in the absence of fluctuations at $Q_{0,s}$, the quartic coefficient $\gamma_{LO}$ diverges as $(b^2-(Q_{0,s}/2)^2)^{-3/2}$ as the field approaches $b_{c,s}$ (see Eq.~\eqref{gamma s Zero T}). Hence $\Delta C/T_c=0$ at zero temperature irrespective of the value of $\eta$.

%%%%%%%%%%%%%%%%%%%%%%%%%%%%%%%%%
%%%%%%%%%%%%%%%%%%%%%%%%%%%%%%%%%%

In contrast, we find that at zero temperature the optimal wave vector for the $d$-wave gap is
\begin{equation}
  Q_{0,d}=e^{-\gamma_E}\exp\left[\frac{\sqrt{3}-1}{4}\right]\xi_0^{-1}
  \approx 0.67\xi_0^{-1}\,,
\end{equation}
with $b_{c,d}/(2Q_{0,d}\xi_0)=(1+\sqrt{3})/4)^{1/2}\approx 0.83$, and the coefficient $\gamma_{LO}(T=0)=0.07$ remains finite for all values of $\eta$. The vanishing of the specific heat jump in the absence of magnetic fluctuations is now due to the vanishing of $\alpha^\prime$ at $T=0$ (discussed in Appendix~\ref{app:numerical Zero T}). The temperature slope of the quadratic term,
$\bar{\alpha}^{\prime}(0)=\alpha^\prime_{LO}(0)+\tEta\chi(0)$, increases as $\eta$ becomes finite, and this leads to a finite
value of $\Delta C/T_c$ for $d$ wave order in the limit
$T\rightarrow 0$ in the presence of the fluctuations.

The negative slope at $t=0$ of the specific heat jump for $d$-wave (Fig.~\ref{fig:Spec Heat}(b)), is due to the temperature dependence of $\chi$. To explain this, we expand Eq.~(\ref{spec heat wrt BCS}) in $t$ to find
\begin{equation}\label{spec heat zero T}
\frac{\Delta C/T_c}{\Delta C/T_{c0}}
\simeq\frac{7\zeta(3)}{8\pi^2}\left(
\frac{(\bar{\alpha}^\prime)^2}{2\bar{\gamma}}+
\frac{\bar{\alpha}^\prime\left(2\bar{\gamma}\bar{\alpha}^{\prime\prime}-\bar{\gamma}^\prime\bar{\alpha}^\prime\right)}{2\bar{\gamma}^2}t\right)
\end{equation}
where all the derivatives and $\bar{\gamma}$ are evaluated at $t=0$.

%%%%%%%%%%%%%%%%%%%%%%%%%%%%%%%%%
\begin{figure}[t]
\includegraphics[width=0.45\textwidth,clip=true]{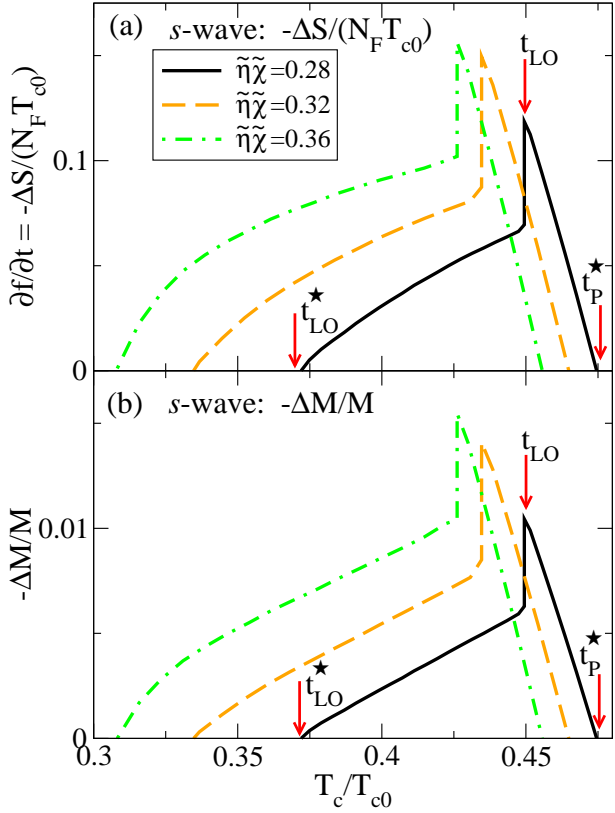}%
%\vspace{-4mm}
\caption{(Color online.) Thermodynamics at first order N-SC
transition for $s$-wave with $\tEta\tChi=$0.28, 0.32, and 0.36. Shown are (a) decrease in entropy and (b) decrease in susceptibility. Temperatures $t_{P}^\star$, $t_{LO}$, and $t_{LO}^\star$ are labeled for $\tEta\tChi=0.28$.}
\label{fig:Entropy DeltaChi}
\end{figure}
%%%%%%%%%%%%%%%%%%%%%%%%%%%%%%%%%%

As discussed in Appendix~\ref{app:numerical Zero T}, for $d$-wave symmetry the quadratic derivative $\bar{\alpha}^{\prime}(0)=\tEta\chi(0)$ is positive while the quartic derivative $\bar{\gamma}^{\prime}(0)=-(\tEta\chi(0))^2/(6t_F)$ is negative at low $t$. The second order quadratic derivative is
$\bar{\alpha}^{\prime\prime}(0)=\alpha_{LO}^{\prime\prime}(0)+2\tEta\chi^{\prime}(0)$ with $\alpha_{LO}^{\prime\prime}(0)\approx 4.54$ and $\chi^\prime=-\tChi/t_{sf}^2$ is always negative for $\chi(T)$. Hence, with $\chi(T)$, the initial slope at $t=0$ of the specific heat jump is determined by how strongly the fluctuations compete with superconductivity. As shown for $\tEta\tChi=0.3$ in Fig.~\ref{fig:Spec Heat}(b), moderate coupling is sufficient to make prominent the dip in the specific heat jump for $d$-wave at low temperatures. For constant susceptibility, however, $\chi^\prime=0$, and the specific heat jump always increases from its value at $t=0$.

\subsubsection{Entropy and magnetization at the first order transition}

Between $t_P^\star$ and $t_{LO}^\star$, where the transition is first order, we compute the entropy jump, $\Delta S=-\partial f/\partial t$, at the
transition, and show it in Fig.~\ref{fig:Entropy DeltaChi}(a).
From Eq.~\eqref{f gen form}, the entropy jump is
\begin{equation}
\begin{split}
-\Delta S
&=\left[\frac{\partial \alpha}{\partial
t}|\delta_0|^2+\frac{\partial \gamma}{\partial
t}|\delta_0|^4+\frac{\partial \nu}{\partial
t}|\delta_0|^6\right]_{t=t_c}\\
&= -\frac{\partial \alpha}{\partial t}
\frac{\gamma}{2\nu}
+\frac{\partial \gamma}{\partial t}
\left(\frac{\gamma}{2\nu}\right)^2
-\frac{\partial \nu}{\partial t}
\left(\frac{\gamma}{2\nu}\right)^3\,
\end{split}
\end{equation}
with $|\delta_0|^2=-\gamma/(2\nu)$ at the first order transition. As the effective coupling parameter between magnetism and superconductivity, $\tEta\tChi$, grows, more and more entropy is transferred at $T_{c}$ from the magnetic fluctuations to superconductivity, and the entropy jump increases. We find, as expected, that $\Delta S$ is largest in the vicinity of $t_{LO}$, where $\delta_0$ takes its maximum value, and is on the order of a few percent of the entropy difference between the SC state at $T=0$ and the normal state at $T_c(B=0)$. We also find that $\delta_0$ is moderate at the first order transition, with its largest value $\delta_0(t_{LO})\approx0.3\delta_0(t=0,b=0)$, and the results of our small $\delta_0$ expansion make physical sense.

The mismatch in the entropy jump in Fig.~\ref{fig:Entropy DeltaChi}(a) at $t_{LO}$ results from
averaging the LO gap amplitude over the system size in the limit $q=0$. Near $t_{LO}$, the wavelength $\lambda_{FFLO} = \pi\xi_0/q$ of $\Delta(x)$ becomes comparable to the system size, and the profile of the order parameter near $t_{LO}$ resembles a single
kink~\cite{BurkhardtFermiLiquidFFLO:1994,AntonFFLO2D:2005} profile that describes the uniform-modulated transition within the SC phase. Below $t_{LO}$ the modulation vector $q_0$ rises rapidly along $b_c(t)$, and the spatial averaging of the order parameter is justified away from the immediate vicinity of $t_{LO}$. Therefore we expect that a calculation free of the single-mode ansatz, will give a greater entropy jump in modulated state in the immediate vicinity of $T_{LO}$.

Since at the first order transition $\gamma$ changes sign, we
expand this coefficient near $t_{P}^\star$ and $t_{LO}^\star$
along the transition line, $\gamma=g_i(t_c-t_i^\star)$ where $g_i$ is positive (negative) near $t_i=t_{P}^\star$ ($t_{LO}^\star)$. We find that, near the tricritical points,
\begin{equation}
-\Delta S=S_{N}(t_c)-S_{SC}(t_c)  \simeq \left.\frac{-g_i}{2\nu}\frac{\partial
\alpha}{\partial t}\right|_{t_c}\hspace{-2mm}(t_c-t_i^\star)\, ,
\end{equation}
where $S_{N}$ and $S_{SC}$ are the entropy in the normal and SC states, respectively. Hence $-\Delta S$ increases linearly in $t_c-t_i$ as seen in Fig.~\ref{fig:Entropy DeltaChi}(a). This behavior may be tested experimentally in magnetocaloric measurements.

Exactly at the points $t_{P}^\star$ and $t_{LO}^\star$, the
entropy difference between the normal and the superconducting
states is zero. Instead, there is a rapid release of entropy upon lowering the temperature at a fixed field, and entering the SC state. Near $t_{P}^\star$ and $t_{LO}^\star$, both the quadratic and quartic Landau coefficients are small and can be expanded about $t_i^\star$, namely $\alpha =a_i(t-t_i^\star)$ and $\gamma =g_i(t-t_i^\star)$. We then find
\begin{equation}
|\delta_0|^2= \frac{-g_i(t-t_i^\star)\pm
\sqrt{g_i^2(t-t_i^\star)^2-3a_i\nu(t-t_i^\star)}}{3\nu}
\end{equation}
and, sufficiently close to $t_i^\star$, $|\delta_0|^2$ is
dominated by the temperature dependence of second term under the square root.  Thus, the entropy relative to the normal state varies with temperature as
\begin{equation}
S_{SC}(t)-S_{N}(t) \simeq
-\frac{3}{2}\sqrt{\frac{a_i^3}{3\nu}}\sqrt{t_i^\star-t}
\end{equation}
where $S_{SC}(t)$ and $S_{N}(t)$ are the entropy in the SC and N states, respectively.

To further test the validity of our parameter choices, we calculate
\begin{equation}
M/H=\left(\chi^{-1}+\eta|\Delta_0|^2_{ave}\right)^{-1}\,,
\end{equation}
along the first order N-SC transition. Here $|\Delta_0|^2_{ave}$ is the spatial average of the SC order parameter. We find that magnetization is suppressed by the onset of superconducting order (see Eq.~\eqref{Fsc,m}) as entropy is transferred between their respective degrees of freedom. The fractional change in magnetization is
\begin{equation}
\frac{\Delta M}{M} =-\frac{\eta\chi(T)|\Delta_0|^2_{ave}}{1+\eta\chi(T)|\Delta_0|^2_{ave}}
\end{equation}
across the transition. This jump, as shown in Fig.~\ref{fig:Entropy DeltaChi}(b), resembles the entropy jump in Fig.~\ref{fig:Entropy DeltaChi}(a), which makes sense as both quantities depend on the value $|\Delta_0|$ takes upon entering the SC state. Thus, the jump in $\delta_0$ across the transition may be revealed by measuring both $\Delta M/M$ and $\Delta S$ along the line of first order transition.

Since we include fluctuations phenomenologically, it is possible that the first order transitions are due to an unreasonable choice of the coupling parameter $\eta$ such that the magnetization is strongly renormalized. As a check on the validity of our model, we verify that the magnetization does not change drastically at the N-SC transition.  As shown in Fig.~\ref{fig:Entropy DeltaChi}(b), the relative change in $M/H$ at a first order transition is generally less than a few percent and validates our method of including of the magnetic fluctuations.

%%%%%%%%%%%%%%%%%%%%%%%%%%%%%%%%%%%%%%%%%%%%%%%%%%%%%%%%%%%%%%%%%%%%%%%%%
\section{Conclusions}

Motivated in part by experiments on the 115 heavy fermion
compounds we considered the effect of a Pauli-limiting Zeeman field on $s$- and $d$- wave superconductors in the presence of classical magnetic fluctuations. We considered both the uniform and inhomogeneous (FFLO) superconducting states, found that the amplitude-modulated state with the modulation vector along the gap nodes is favorable in the $d$-wave case, and investigated the order of the transition. Our main finding is that there exists a range of temperatures, in the vicinity of the onset of the modulated state, where the normal metal-to-superconductor transition is first order \emph{both into a uniform and into a modulated state.} The width of the temperature range increases with the strength of coupling to the magnetic fluctuation and is generally greater for $s$-wave systems.

While we considered only classical thermal fluctuations, since the regime of interest occurs for temperatures $T/T_c\sim 0.2-0.5$ we believe that this approach is sufficient. The question of whether the transition remains second order as $T\rightarrow 0$, \emph{e.g.}, when accounting for the quantum dynamics of spins, is left for future studies. Among other potentially interesting issues are whether impurity scattering, which is known to suppress the inhomogeneous LO state, enhances or shrinks the first order transition regime, what the results of combining the Zeeman field with the orbital coupling and vortex physics would be. In the present form our approach outlines a new, generic, path towards a first order N-SC transition, and demonstrates one experimentally observed feature: the separation between the onset of the first order transition and the transition into a modulated state. It suggests that accounting for magnetic fluctuations which are known to exist in heavy fermion and other related compounds affects the shape of  the transition lines, the order of the transition, and the behavior of the thermodynamic properties at the transition.

%%%%%%%%%%%%%%%%%%%  Acknowledgements %%%%%%%%%%%%%%%%%%%%%%%
\begin{acknowledgments}
This work was supported in part by the Louisiana Board of Regents
and by US DOE via Grant No. DE-FG02-08ER46492.
\end{acknowledgments}

%%%%%%%%%%%%%%%%%%%%%%  Appendix  %%%%%%%%%%%%%%%%%%%%%%%%%%%%
\appendix

%%%%%%%%%%%%%%%%  Expansion in Delta  %%%%%%%%%%%%%%%%%%%%%%%%
\section{Expansion in $\Delta_\vq$}\label{GLF derivation}

To derive the coefficients for the Landau free energy functional, we begin with the requirement $\cF=-T \ln(\cZ)$ is an extremum with respect to $\Delta_\vq$ and
$\Delta^*_\vq$, so that
\begin{equation}\label{vary F full}
\delta\cF=
\sum_\vq\left(\frac{\Delta^*_\vq}{|\lambda|}+\sum_k\cY(\hat{\vk})
\left<c_{-\vk,-}^\dagger
c_{\vk+\vq,+}^\dagger\right>\right)\delta\Delta_\vq
+ h.c. \,.
\end{equation}
We construct the Landau free energy functional by expanding in powers of $\Delta_{\vq}$, but not in the modulation wave vector $\vq$, which allows us to treat the low temperature region. To carry out this expansion we use the Gor'kov formulation of the Green's function approach. The normal,
\begin{equation} G_\sigma(\vk,\vk^\prime; \tau) =
-\left<T_\tau\left(c_{\vk,\sigma}(\tau)c^\dagger_{\vk^\prime,\sigma}(0)\right)\right>,
\end{equation}
and anomalous,
\begin{equation}
F^\dagger(\vk,\vk^\prime; \tau) =
-\left<T_\tau\left(c^\dagger_{\vk,-}(\tau)c^\dagger_{\vk^\prime,+}(0)\right)\right>,
\end{equation}
Green's function satisfy
\begin{eqnarray}
\nonumber &&(i\omega_n
-\epsilon_{\mathbf{k,+}})G_+(\vk,\mathbf{k^\prime};i\omega_n)\\
&& +\sum_{\vq}\cY(\hat{\vk})\Delta_\vq
F^\dagger(-\vk+\vq,\vk^\prime;i\omega_n)=\delta_{\vk,\vk^\prime}\,,
\label{gorkov G}
\\
\nonumber && (i\omega_n
+\epsilon_{-\mathbf{k,-}})F^\dagger(-\vk,\mathbf{k^\prime};i\omega_n)\\
&&+\sum_{\vq}\cY(\hat{\vk})\Delta^*_\vq
G_+(\vk+\vq,\vk^\prime;i\omega_n) =0\,,
\label{gorkov F}%\end{split}
\end{eqnarray}
respectively. Here $T_\tau$ denotes imaginary time ordering, and $\omega_n=2{\pi}T\left(n+\frac{1}{2}\right)$ is the fermionic Matsubara frequency. The thermal average entering the free energy expression, Eq.~(\ref{vary F full}), is given by
\begin{equation}\label{thermal average}
\left<c_{-\vk,-}^\dagger
c_{\vk+\vq,+}^\dagger\right>
=-T\hspace{-0.5mm}\sum_{n}\hspace{-0.5mm}F^\dagger(-\vk,\vk+\vq;i\omega_n)\,.
\end{equation}

We iteratively expand Eqs.~\eqref{gorkov G} and \eqref{gorkov F} in powers of $\Delta_\vq$ and $\Delta_\vq^*$, and hence find the series expansion for  $F^\dagger(-\vk,\vk^\prime;i\omega_n)$.  Using this expansion for the thermal average in Eq.~\eqref{vary F full}, we integrate term by term with respect to $\Delta_\vq$, and we obtain the Landau free energy density, $F_{L}=\cF/L^2$ where $L^2$ is the 2D system size, up to $\cO(|\Delta_0|^6)$ inclusive. We find
\begin{equation} \label{gen glf appendix}
\begin{split}
    F_{L} &= \sum_{ \{\vq_i\}}\widetilde{\alpha}_{\vq_i} |\Delta_{\vq_i}|^2 \\
    & +\sum_{ \{\vq_i\}}\widetilde{\gamma}_{\vq_1,\ldots,\vq_4}\Delta_{\vq_1}\Delta^*_{\vq_2}\Delta_{\vq_3}\Delta^*_{\vq_4}\delta_{\vq_1+\vq_3,\vq_2+\vq_4}\\
    & +\sum_{ \{\vq_i\}}\widetilde{\nu}_{\vq_1,\ldots,\vq_6}\Delta_{\vq_1}\Delta^*_{\vq_2}\Delta_{\vq_3}\Delta^*_{\vq_4}\Delta_{\vq_5}\Delta^*_{\vq_6}\\
    & \hspace{2cm}\times \delta_{\vq_1+\vq_3+\vq_5,\vq_2+\vq_4+\vq_6}\,
\end{split}
\end{equation}
where the summation over $\{\vq_i\}$ includes all possible
combinations of the allowed Fourier components of $\Delta(\vr)$.

The fully $\vq$-dependent coefficients of the Ginzburg-Landau
expansion are given by
\begin{equation}\label{alpha general}
\begin{split}
\widetilde{\alpha}_{\vq}
=\frac{1}{|\lambda|}-T\sum_{n,\vk}
|\cY(\hat{\vk})|^2&
G^0_+(\vk+\vq;i\omega_n)\\
&\times G^0_-(-\vk;-i\omega_n),
\end{split}
\end{equation}
\begin{equation}\label{gamma general}
\begin{split}
\widetilde{\gamma}_{\vq_1,...,\vq_4} =
\frac{T}{2}&\sum_{n,\vk}|\cY(\hat{\vk})|^4 G^0_+(\vk+\vq_1;i\omega_n)\\
&\times G^0_-(-\vk+\vq_3-\vq_2;-i\omega_n)\\
&\times
G^0_+(\vk+\vq_2;i\omega_n)G^0_-(-\vk;-i\omega_n),
\end{split}
\end{equation}
\begin{equation}\label{nu general}
\begin{split}
\widetilde{\nu}_{\vq_1,...,\vq_6} =&-
\frac{T}{3}\sum_{n,\vk}|\cY(\hat{\vk})|^6 G^0_+(\vk+\vq_1;i\omega_n)\\
&\times G^0_-(-\vk+\vq_3+\vq_5-\vq_2-\vq_4;-i\omega_n)\\
&\times  G^0_+(\vk+\vq_2+\vq_4-\vq_3;i\omega_n)\\
&\times G^0_-(-\vk+\vq_3-\vq_2;-i\omega_n)\\
&\times
G^0_+(\vk+\vq_2;i\omega_n)G^0_{-}(-\vk;-i\omega_n),
\end{split}
\end{equation}
where
\begin{equation}\label{g0 appendix}
G^0_\pm(\vk;i\omega_n) = \left[ i \omega_n
-\epsilon_\mathbf{k,\pm}\right]^{-1}
\end{equation}
is the normal state propagator for an electron  in a Zeeman field. After integration over $\vk$, the interaction strength $|\lambda|$ in the quadratic coefficient $\alpha$ will be eliminated in favor of the zero-field transition temperature $T_{c0}$. Assuming a circular Fermi surface, we use 2D angular basis functions
\begin{equation}\label{basis fxns}
\cY(\hat{\vk}) =
\begin{cases}
1 & s\rm{-wave} \\
\sqrt{2}\left(\hat{k}_x^2-\hat{k}_y^2\right) = \sqrt{2}\cos(2\theta)
& d_{x^2-y^2},
\end{cases}
\end{equation}
normalized so that $\left<|\cY(\theta)|^2\right>_{FS} = 1$
where $\theta$ is the azimuthal angle in momentum space. Here
$\left<\cdots\right>_{FS}$ indicates an average over the 2D Fermi surface. All the momentum sums are evaluated using the fact the the Green's functions are peaked at the Fermi energy, so that, for our model of a 2D circular Fermi surface
\begin{equation}\label{sum to int}
\sum_\vk\rightarrow
\frac{N_F}{2\pi}\int_0^{2\pi}d\theta\int_{-\infty}^\infty
d\epsilon\,.
\end{equation}

%%%%%%%%%%%%%%%%%  Tridiagonal Integral %%%%%%%%%%%%%%%%%%%%%
\section{Tridiagonal Integral} \label{app:TriDiagInt}

For the case of single-mode $\cos(\vQ\cdot\vr)$ modulation of the order parameter, the magnetic contribution to the free energy functional (due to the off-diagonal $\vk,\vk\pm2\vQ$ coupling) takes the tridiagonal form
\begin{equation}\label{tridiag funct}
\begin{split}
\cF(\vM(\vr)) =& T\sum_\vk\left(a_\vk
\vM_\vk\cdot\vM^\star_\vk\right.\\
&+\left.b_\vk\vM_\vk
\cdot\left(\vM^\star_{\vk+2\vQ}+\vM^\star_{\vk-2\vQ}\right)\right)
\end{split}
\end{equation}
where
\begin{equation}\label{ak}
a_\vk = a \equiv
\frac{1}{2T}\left(\frac{1}{\chi}+\frac{1}{2}\eta|\Delta_0|^2\right),\
\ \forall \vk
\end{equation}
and
\begin{equation}\label{bk}
b_\vk = b \equiv -\frac{1}{8T}\eta|\Delta_0|^2, \ \ \forall
\vk.
\end{equation}
This yields the partition sum
\begin{equation}\label{mod partition sum}
\begin{split}
\cZ &=\prod_\vk\int \cD(\vM_\vk) \exp\left[-\left(a |\vM_\vk|^2\right.\right.\\
&\hspace{1.2cm}+ \left.\left.b\vM_\vk\cdot\left(\vM^\star_{\vk+2\vQ}+\vM^\star_{\vk-2\vQ}\right)\right)\right]\\
&=\prod_{i=1}^d\prod_\vk\int \cD (M_{\vk,i})\exp\left[-\left(a |M_{\vk,i}|^2\right.\right.\\
&\hspace{1.2cm}+\left.\left. b
M_{\vk,i}\left(M^\star_{{\vk+2\vQ},i}+M^\star_{{\vk-2\vQ},i}\right)\right)\right]
\end{split}
\end{equation}
where the product over $i$ accounts for the $d$ spatial components of $\vM(\vr)$.  To compute this integral, we separate the product over all wave vectors into a product over components parallel and perpendicular to the direction of $\vQ$.  As the terms comprising $\cZ$ have no functional dependence on $i$, we have $\cZ=\cZ_0^d$ where
\begin{equation}
\begin{split}
\cZ_0 &=\prod_{k_\perp}\prod_{k_\parallel}\int
\cD(M_{k_\perp,k_\parallel}) \exp\left[-a |M_{{k_\perp,k_\parallel}}|^2\right]\\
&\times\exp\left[ b M_{k_\perp,k_\parallel}
\left(M^\star_{k_\perp,k_\parallel+2Q}+M^\star_{k_\perp,k_\parallel-2Q}\right)\right]\,.
\end{split}
\end{equation}

Due to the coupling between $M_{k_\perp,k_\parallel}$ and
$M_{k_\perp,k_\parallel \pm 2Q}$, the product over $k_\parallel$ can be divided up into a product of integrals taken only over wave vectors $|k_\parallel|\le|Q|$, effectively employing the Brillouin zone method of solid state physics with $|k_\parallel|\le|Q|$ corresponding to the first Brillouin zone. Each term in the product over $|k_\parallel|\le|Q|$ is then an integral connecting $k_\parallel$ to $k_n=k_\parallel + n(2Q)$, where $n$ is an integer. The sum over $k$ is cut off at a wave vector on the order of the inverse lattice spacing $k_c = \pi/l$.  So, to cut off the sum over $n$, we define the cut off integer $n_c$ such that $k_{\pm n_c}=k_\parallel\pm n_c (2Q)\approx k_c$.

Separating the product over $k_\parallel$ in this way, and
introducing the notational shorthand \[M_n(k_\perp,k_\parallel)=M_{k_\perp,k_\parallel+n(2Q)}\,,\] our partition sum can be rewritten
\begin{equation}\label{expanded part sum}
\begin{split}
&\cZ_0
=\prod_{k_\perp}\prod_{k_\parallel=-Q}^Q\int\left[\cdots\cD(M_{1})\cD(M_{-1})\cD(M_0)
\right]\\
&\hspace{4mm}\times\exp\left[-a |M_0|^2 - b M_0\left(M^\star_{1} +
M^\star_{-1}\right)-\cdots\right].
\end{split}
\end{equation}
We integrate over the real and imaginary parts of $M_n = M'_n+iM''_n$ and restrict the product over $\vk$ to be over one-half of $k$-space because $M_\vk=M_{-\vk}^\star$ for real $\vM(\vr)$.  However, as the integrand factors into two identical integrals over $M'_n$ and $M''_n$, we can integrate over $M'_n$ alone and take the product over all values of $\vk$. Thus
\begin{equation}\label{expanded part sum 2}
\begin{split}
&\cZ_0
=\prod_{k_\perp}\prod_{k_\parallel=-Q}^Q\int\left[\cdots\cD(M'_{1})\cD(M'_{-1})\cD(M'_0)
\right]\\
&\hspace{4mm}\times\exp\left[-a (M'_0)^2 - 2 b M'_0\left(M'_{1} +
M'_{-1}\right)-\cdots\right].
\end{split}
\end{equation}

Beginning with $n=0$, we integrate recursively over all $M'_n$ and denote by $a_n$ and $b_n$ the renormalized coefficients of $(M'_{\pm n})^2$ and $M'_nM'_{-n}$, respectively.  Working with $a$ and $b$ given in Eqs.~\eqref{ak} and \eqref{bk}, the integration coefficients are
\begin{equation}\label{coeff 0}
a_{0} =a \hspace{2mm}\mathrm{and}\hspace{2mm} b_{0} =b
\hspace{2mm}\mathrm{for}\hspace{2mm}n=0,
\end{equation}
\begin{equation}\label{coeff 1}
a_{1} =a_0-\frac{b_0^2}{a_0} \hspace{2mm}\mathrm{and}\hspace{2mm}
b_{1} =\frac{b_0^2}{a_0} \hspace{2mm}\mathrm{for}\hspace{2mm}n=1,
\end{equation}
and, for $n\ge1$, the remaining terms
\begin{equation}\label{coeff n}
a_{n+1} =a_0-\frac{ a_{n}b_0^2}{a_{n}^2 -b_{n}^2}
\hspace{2mm}\mathrm{and}\hspace{2mm} b_{n+1} =\frac{b_{n}b_0^2}{a_{n}^2 -b_{n} ^2}.
\end{equation}
are determined recursively. The partition sum becomes
\begin{equation}\label{part sum product}
\begin{split}
\cZ_0
&=\prod_{k_\perp,k_\parallel}\sqrt{\frac{\pi}{a_0}}
\sqrt{\frac{\pi^2}{a_{1}^2
-b_{1}^2}}\times\cdots\times\sqrt{\frac{\pi^2}{a_{n_c}^2 -b_{n_c}^2}}\\
&=\prod_{k_\perp,k_\parallel}\left(\sqrt{\frac{\pi}{a_0}}\right)^{2n_c+1}
\sqrt{\frac{a_0^2}{a_{1}^2 -b_{0}^2}}\times\cdots
\end{split}
\end{equation}
where $k_\parallel\in(-Q,Q)$ is understood. The free energy functional $\cF(\vM(\vr))$ can now be replaced with its thermodynamic average $\cF=-\beta^{-1}\ln\left(\cZ\right)$ which is
\begin{equation}\label{sum n.gt.0}
\cF
=\frac{d}{2\beta}\left[\sum_{|\vk|=0}^{k_c}\ln\left(\frac{a_0}{\pi}\right)+\sum_{\substack{k_\perp,k_\parallel\\n}}^{\prime}\ln\left(
\frac{a_{n}^2 -b_{n}^2}{a_0^2}\right)\right]\,,
\end{equation}
where for the second sum $n\in(-n_c,n_c)$. The prime implies that $n=0$ is excluded from the sum since the $n=0$ term is $\ln(1)=0$.

In order to obtain the necessary small $\Delta$ expansion of Eq.~\eqref{sum n.gt.0}, we need to expand $\left(a_{n}^2- b_{n}^2\right)/a_0^2$ to $\cO(|\Delta_0|^6)$ inclusive. We do this by introducing recursion relations
\begin{equation}
\begin{split}
s_n&=a_n+b_n=a_0-\frac{b_0^2}{s_n}\,,\quad n>1\\
d_n&=a_n-b_n= a_0-\frac{b_0^2}{d_n}\,,\quad n>1
\end{split}
\end{equation}
with the initial values $s_1=a_0$ and $d_1=a_0-2b_0^2/a_0$, respectively. Taking $a_0=a$ and $b_0=b$ from Eqs.~\eqref{ak} and \eqref{bk}, we expand $s_nd_n/a_0^2=\left(a_{n}^2-b_{n}^2\right)/a_0^2$ to third order in $b$ since $b\propto|\Delta_0|^2$.  Expressing $a$ and $b$ in units of $1/2\chi T$, we have the initial values
\begin{equation}
\begin{split}
s_1&=a=1-2b\\ d_1&=a-\frac{2b^2}{1-2b}=1-2b-2b^2-4b^3+\cO(b^4)\,,
\end{split}
\end{equation}
and the remaining terms for $n>1$
\begin{equation}
s_n=d_n=1-2b-b^2-2b^3+\cO(b^4)\,.
\end{equation}
With these expressions, we find that $s_nd_n=1-4b+2b^2$ is independent of $n$ when expanded to third order in $b$. Thus,
\begin{equation}\label{ratio s and d} \frac{s_nd_n}{a_0^2}=\frac{1-4b+2b^2}{(1-2b)^2}=1-2b^2-8b^3+\cO(b^4)\,,
\end{equation}
and, substituting $b=-\eta\chi|\Delta_0|^2/4$, we finally obtain
\begin{equation}\label{ratio a and b}
\frac{a_{n}^2- b_{n}^2}{a_0^2}
=1-\frac{1}{8}\eta^2\chi^2|\Delta_0|^4+\frac{1}{8}\eta^3\chi^3|\Delta_0|^6.
\end{equation}
up to $\cO(|\Delta_0|^6)$ inclusive. Since the summands no longer depends on $n$, we recollect the summation over $k_\perp$, $k_\parallel$, and $n$ into a sum over $|\vk|<k_c$. We take the sum to include all $n\in(-n_c,n_c)$ with the $n=0$ term identical to the rest. We justify this by noting that, for a system of size $L^D$, the sum over $k_\parallel$ for $n=0$ is of order $2QL$ and is much smaller than the sum over all $k_\parallel<k_c$ (of order $2k_cL$) since $Q\ll k_c$ (where $Q\lesssim\xi_0^{-1}$ and $k_c = \pi/l$).

After subtracting the average magnetic contribution to the normal state, the fluctuation contribution to the superconducting free energy is
\begin{equation}
\begin{split}
\cF_{LO,M}
&=\frac{d}{2\beta}\sum_{|\vk|=0}^{k_c}\left[\ln\left(1+\frac{1}{2}\eta\chi|\Delta_0|^2\right)\right.\\
&\hspace{0.0cm}\left.+\ln\left(1-\frac{1}{8}\eta^2\chi^2|\Delta_0|^4+\frac{1}{8}\eta^3\chi^3|\Delta_0|^6\right)\right],
\end{split}
\end{equation}
which, with $d=3$, is the expression given in Eq.~\eqref{Fmod
flucts}.

%%%%%%%%%%%%%%%  Zero temperature limit  %%%%%%%%%%%%%%%%%%%%
\section{Zero temperature limit} \label{app:Zero T limit}

We determine the Landau coefficients $\bar{\alpha}_{LO}$ and $\bar{\gamma}_{LO}$ and their temperature derivatives in the limit $t\rightarrow 0$.  We first derive analytically $\alpha_{LO}$ and $\gamma_{LO}$ from Eqs.~\eqref{alpha FF bare} (with prefactor of $1/2$ for LO) and \eqref{gamma LO bare} for both $s$- and $d$-wave at zero temperature; there the magnetic fluctuations die out so that $\bar{\alpha}_{LO}=\alpha_{LO}$ and $\bar{\gamma}_{LO}=\gamma_{LO}$. We then determine numerically the derivatives $\alpha_{LO}^\prime(0)$, $\alpha_{LO}^{\prime\prime}(0)$, and $\gamma_{LO}^\prime(0)$ for $d$-wave symmetry and add to them the magnetic fluctuation corrections at $t=0$.

\subsection{Analytic determination of $b_c$, $q_0$, and $\gamma_{LO}$}\label{app:analytic Zero T}

We first evaluate the quadratic Landau coefficient for the LO gap modulation.  In the limit $t=0$, the quadratic coefficient becomes
\begin{equation}\label{alpha LO Zero T}
\alpha_{LO}=\frac{1}{4}\left<|\cY(\theta)|^2 \ln\left(\left(b+\bq\right)^2\right)\right>-\frac{1}{2}\Psi\hspace{-1mm}\left(\frac{1}{2}\right)\,,
\end{equation}
where $\cY(\theta)=1$ and  $\cY(\theta)=\sqrt 2 \cos 2\theta$ for $s$- and $d$-wave gaps, respectively. Here we use $b=\mu B/(2\pi T_{c0})$ and $\bar{q}=q \cos(\theta-\theta_q)$ where $q =\xi_0 Q/2$. The angle $\theta_q$ is the modulation direction with respect to the crystalline $a$ axis, and $\theta_q=\pi/4$ for nodally-oriented $d$-wave. Integration over $\theta$ yields
\begin{equation}\label{alpha s Zero T}
\alpha_{LO,s}=\frac{1}{2}\re\left[\ln\left(\frac{b+\sqrt{b^2-q^2}}{2}\right)\right]-\frac{1}{2}\Psi\hspace{-1mm}\left(\frac{1}{2}\right)
\end{equation}
and
\begin{equation}\label{alpha d Zero T}
\alpha_{LO,d}=\frac{1}{4}\ln\left(\frac{q^2}{4}\right)+\frac{b^4}{q^4}-\frac{b^2}{q^2}+\frac{1}{8}-\frac{1}{2}\Psi\hspace{-1mm}\left(\frac{1}{2}\right)
\end{equation}
for $s$- and $d$-wave respectively. We locate the transition by finding the maximum $b$ for which $\alpha_{LO,d}=0$, and we
find that for $s$-wave
\begin{equation}\label{q and b Zero T swave}
q_{0,s}=b_{c,s}=2e^{\Psi(1/2)}=\frac{e^{-\gamma_E}}{2}\simeq0.281\,,
\end{equation}
where $\gamma_E\approx0.577$ is Euler's constant, and for $d$-wave
\begin{equation}\label{q and b Zero T dwave}
\begin{split}
q_{0,d}&=\frac{e^{-\gamma_E}}{2}\exp\left(-2a^4+2a^2-\frac{1}{4}\right)\simeq0.337\,,\\ %0.33711
b_{c,d}&=a q_{0,d}\simeq0.278\,,
%0.27860
\end{split}
\end{equation}
where $a=((1+\sqrt{3})/4)^{1/2}\simeq0.826$.

To determine the quartic Landau coefficient, first note that,
in the limit $T=0$, the Matsubara sum $2\pi T \sum_{n}F(\w_n)$ becomes the integral $\int d\w F(\w)$. Thus, we rewrite Eq.~\eqref{gamma LO bare} as
\begin{equation}\label{gamma LO Zero T}
\begin{split}
\gamma_{LO}&=\re\int_{0}^\infty \frac{d\bw}{128 \pi ^2}\int_{0}^{2\pi} \frac{d\theta}{2\pi} |\cY(\theta)|^4 I_\gamma(\bw,b,q,\theta)
\end{split}
\end{equation}
where $\bw = \w/(2\pi T_{c0})$ and
\begin{equation}\label{gamma integrand}
I_\gamma(\bw,b,q,\theta)=
\frac{\left(\bw+ib\right) \left(3 \left(\bw+ib\right)^2-\bq^2\right)}{\left(\left(\bw+Ib\right)^2+\bq^2\right)^3}\,.
\end{equation}

We perform the angular integration changing variables to $z=e^{i\theta}$, and then integrating around the unit circle in the complex $z$-plane. After thus averaging over the Fermi surface, we arrive at
\begin{equation}
I_{\gamma,s}(\bw,b,q)=
\frac{2q^4+5q^2\bw_b^2+6\bw_b^4}{2\bw_b^2(q^2+\bw_b^2)^{5/2}}
\end{equation}
and
\begin{equation}
\begin{split}
I_{\gamma,d}(\bw,b,q)
&=
\frac{24\bw_b}{q^8}\frac{8q^4\bw_b+44q^2\bw_b^3+40\bw_b^5}{\sqrt{q^2+\bw_b^2}}\\
&\hspace{5mm}-\frac{24\bw_b}{q^8}\left(q^4+24q^2\bw_b^2+40\bw_b^4\right)
\end{split}
\end{equation}
for $s$- and $d$-wave respectively. Here $\bw_b=\bw+i b$.  Evaluating the integral over $\bw$ we arrive at
\begin{equation}\label{gamma s Zero T}
\gamma_{LO,s}=\frac{3}{32\pi^2}\frac{3b^2-2q^2}{b(b^2-q^2)^{3/2}}\,,
\end{equation}
and
\begin{equation}\label{gamma d Zero T}
\gamma_{LO,d}=\frac{1}{64\pi^2q^2}\left(1-2\frac{b^2} {q^2}\left(3-36\frac{b^2}{q^2}+40\frac{b^4}{q^4}\right)\right)
\end{equation}
the quartic Landau coefficients at $t=0$. From Eq.~\eqref{gamma s Zero T}, we find that $\gamma_{LO,s}$ diverges as $b_c\rightarrow q_0$ (see Eq.~\eqref{q and b Zero T swave}) while, from Eq.~\eqref{q and b Zero T dwave} and \eqref{gamma d Zero T} we see that $\gamma_{LO,d}\simeq0.070$ remains finite when $T\rightarrow0$.

\subsection{Evaluation of derivatives for $d$-wave at $T=0$}
\label{app:numerical Zero T}

The temperature derivatives of the quadratic and quartic coefficients are
\begin{subequations}
\begin{eqnarray}
\bar{\alpha}_{LO}^\prime(\tEta,t)&=&\alpha_{LO}^\prime(t)+\frac{1}{2}\tEta\chi\\
\bar{\alpha}_{LO}^{\prime\prime}(\tEta,t)&=&\alpha_{LO}^{\prime\prime}(t)+\tEta\chi^\prime\\
\bar{\gamma}_{LO}^\prime(\tEta,t)&=&\gamma_{LO}^\prime(t)-\frac{1}{6t_F}\tEta^2\chi^2
\end{eqnarray}
\end{subequations}
Expressions for $\alpha_{LO}^\prime(t)$, $\alpha_{LO}^{\prime\prime}(t)$, and $\gamma_{LO}^\prime(t)$ are obtained by taking the first and second derivatives of Eq.~\eqref{alpha FF bare} (with prefactor of $1/2$ for LO modulation) and the first derivative of Eq.~\eqref{gamma LO bare}, respectively, with respect to $t$.

\begin{figure}[t]
\includegraphics[width=0.45\textwidth,clip=true]{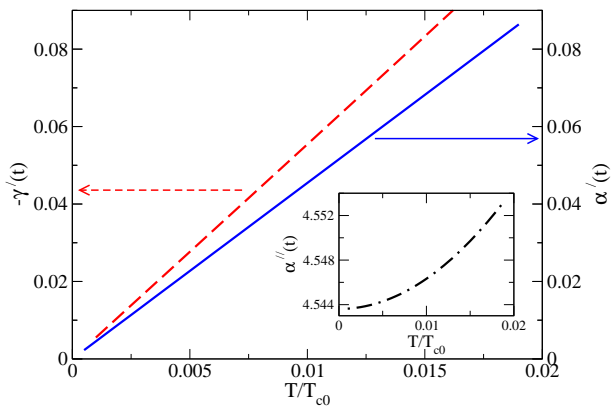}%
\caption{(Color online.) Temperature derivatives of quadratic and quartic Landau coefficients for $d$-wave at fixed  $b_{c,d}$ and $q_{0,d}$ in zero temperature limit.  Main figure shows $\alpha_{LO}^\prime$ and $-\gamma_{LO}^\prime$, both of which limit to zero at $t=0$. Inset: $\alpha_{LO}^{\prime\prime}\approx4.54$ at $t=0$.}
\label{fig:Zero Temp Inset}
\end{figure}

We determine $\alpha_{LO}^\prime(0)$, $\alpha_{LO}^{\prime\prime}(0)$, and $\gamma_{LO}^\prime(0)$ numerically by fixing $b=b_{c,d}$ and $q=q_{0,d}$ and evaluating the derivatives as $t$ approaches zero. As shown in Fig.~\ref{fig:Zero Temp Inset}, we find that $\alpha_{LO}^\prime=\gamma_{LO}^\prime=0$ and  $\alpha_{LO}^{\prime\prime}\approx 4.544$ at $t=0$. Using these values and working with $\chi(T)$, we obtain
$\bar{\alpha}_{LO}^\prime(0)=6.18\tEta\tChi$, $\bar{\alpha}_{LO}^{\prime\prime}(0)=4.544-152.8\tEta\tChi$, and $\bar{\gamma}_{LO}^\prime(0)=-1.48(\tEta\tChi)^2$
in the zero temperature limit.

%%%%%%%%%%%%%%%%%%%%%%%%%%%%%%%%%
%%%%%%%%%%%%%%%%%%%%%%%%%%%%%%%%%%

%\bibliographystyle{apsrev}

%\bibliography{MagFluctsRefs}

\end{document}